# An automatic approach to explore multi-reaction mechanism for medium-sized bimolecular reactions *via* collision dynamics simulations and transition state searches

*Qinghai Cui[1], Jiawei Peng[1], Chao Xu[1,2], Zhenggang Lan[1,\*]*

[1] Guangdong Provincial Key Laboratory of Chemical Pollution and Environmental Safety and MOE Key Laboratory of Environmental Theoretical Chemistry, SCNU Environmental Research Institute, School of Environment, South China Normal University, Guangzhou 510006, P. R. China.

[2] Key Laboratory of Theoretical Chemistry of Environment, Ministry of Education; School of Chemistry, South China Normal University, Guangzhou 510006, P. R. China.

**Abstract**

We develop a broadly-applicable computational method for the automatic exploration of the bimolecular multi-reaction mechanism. The current methodology mainly involves the high-energy Born-Oppenheimer molecular dynamics (BOMD) simulation and the successive reaction pathway construction. Several computational tricks are introduced, which include the selection of the reactive regions based on the electronic-structure calculations and the employment of the virtual collision-dynamics simulations with monitoring atomic distance before BOMD. These prescreening steps largely reduce the number of trajectories in the BOMD simulations and significantly save computational cost. The hidden Markov model combined with modified atomic connectivity matrix is taken for the detection of reaction events in each BOMD trajectory. Starting from several geometries close to reaction events, the further intermediate optimization and transition-state searches are conducted. The proposed method allows us to build the complicated multi-reaction mechanism of medium-sized bimolecular systems automatically. Here we examine the feasibility and efficiency of the current method by its performance in searching the mechanisms of two prototype reactions in environmental science, which are the penicillin G anion + $H_2O$ and the penicillin G anion + OH radical reactions. The result indicates that the proposed theoretical method is a powerful protocol for the automatic searching of the bimolecular reaction mechanisms for medium-sized compounds.

# 1. Introduction

The computational studies on the reaction mechanism provide the crucial understanding of chemical reactions, and the representative progresses of this topic may be found in some review articles.[1-4] At the same time, the studies of reaction mechanism also play essential roles in various chemical-relevant research fields, such as the understanding of chemical conversions of pollutant compounds in environmental sciences[5-8] and the investigation of the biological catalysis processes in biochemistry.[9,10] Thus, the development of the novel user-friendly automatic computational approach to find the reaction mechanism is an essential topic in theoretical chemistry.

One of the center tasks in the computational studies of reaction mechanism is to find transition states (TSs). The successful TS search is highly dependent on the chemical knowledge and intuition of individual researchers, and this brings difficulty for non-experts in practices. It is also possible to employ some automatic TS searching approaches, such as Nudged Elastic Band (NEB),[11] growing-string method (GSM),[12,13] and various string methods.[14-16] However, all above approaches heavily rely on the proper initial information of reactant and product of each elementary reaction, *i.e.,* local minimum-energy structures in reaction pathways, while the generation of these structures may not be an easy task for unknown reactions. Therefore, it is great interesting to develop effective theoretical approaches for the automatic exploration of complicated reaction mechanism.

The development of black-box theoretical approaches for automatic reaction mechanism discovery is a great challenging task in theoretical chemistry, and several approaches were proposed in the last decades, as summarized by available reviews.[2,3,17,18] Among them, some works tried to generate the reaction mechanism by using the knowledge of the structure-reactivity information based on databases or chemical heuristic rules.[19-25] Differently, several groups tried to generate the possible involved species firstly, and then used them as the starting points to perform TS search in the automatic reaction mechanism construction. The possible products may be generated by the graph representation of molecular connectivity, and these species are connected by various string approaches in the automatic reaction pathway generation.[2,26-28] Alternatively, the reactive intermediates may be generated by heuristic-guided approaches and then the constrained optimization may give the initial guesses of the reaction pathways.[29-31] Starting from reactants, it is also possible to employ the reaction-pathway guided TS search in mechanism construction by adding the bias potentials in the trajectory propagation[32-37] or distorting molecules along some critical coordinates (such as the normal-mode coordinates with strongest anharmonicity).[4,38-40]

In theoretical approaches of the automatic discovery of reaction mechanism, one important idea is to use the Born-Oppenheimer molecular dynamics (BOMD). Our proposed approach also heavily relies on such idea. Given reactants, the on-the-fly BOMD simulation should in principle give all possible chemical reactions, once enough energy is added into systems.[41-48] This implies that the reaction mechanism may be directly extracted from the BOMD simulations. According to such idea, Martínez-Núñez and co-workers developed a method called "the transition state search using chemical dynamics simulation (TSSCDS)".[41,42,45,46] The high-energy BOMD is run at the semi-empirical level, which provides the basic information of chemical reactions. When the bond cleavage and formation in BOMD were detected by monitoring the atomic connectivity, the geometries near the reaction events were chosen to perform the TS optimization. The TSSCDS method was effective in the construction of unimolecular reaction network. This approach was extended to study catalysis reactions, in which initial reactant complexes were generated by putting two randomly-orientated reactive species together.[44] Recently, Martínez-Núñez, Shalashilin, Glowacki and their co-workers tried to develop a theoretical approach for the automatic discovery of reaction network by using the boxed molecular dynamics[49-51] in energy space,[47] which is a powerful approach to sample rare events in molecular dynamics. Later, they developed the 'Chemical Network Mapping through combined Dynamics and Master Equation simulations (ChemDyMe)' method.[48] Based on the updated version of the TSSCDS approach, an open source package named AutoMeKin2021[52] was released recently.

When several reactants are involved, the whole reaction network in principle becomes very complicated. To address this challenge, Martínez and co-workers developed an ab initio nanoreactor simulation tool.[53,54] The long-time BOMD including many compounds was run under high temperature and pressure, which allows the generation of various chemical reactions. To manage the huge computational cost, the electronic structure calculations were speeded up using graphics processor units. The hidden Markov model[55] (HMM) analysis of the atomic connectivity matrix was employed to extract reaction events in the BOMD trajectories. The geometries in adjacent to the reaction steps were taken to conduct the intermediate and TS optimization. They also developed the trajectory smoothing method to provide better initial guesses for the reaction pathway construction.[54] The nanoreactor method was further extended by Lei and their coworkers, who used DFTB[56,57] (density functional tight-binding) in BOMD.[58] Alternatively, the chemical dynamics involving many compounds may be efficiently simulated by using ReaxFF[59-61] (a widely-used reactive force field) and even neural-network potentials, and such simulations may produce the very complicated reaction network, as shown by Zhu, Zhang and co-workers.[62,63]

Inspired by the above works of the reaction-mechanism discovery relying on BOMD followed by the TS optimization or reaction pathway construction, we made our initial effort to develop a theoretical method for the automatic exploration of reaction mechanism. In current stage, we focus on bimolecular reactions of medium-sized compounds. Roughly speaking, our approach includes high initial collision-energy BOMD simulations and successive reaction pathway constructions. Several useful technical tricks were applied in our implementation, which largely reduce computational cost and significantly improve optimization convergence. Initially, the reactive regions are selected for each reactant according to the reactivity properties derived from electronic-structure calculations. As we assume that chemical reactions should only happen between chosen reactive regions, the collision dynamics between each pair of reactive centers are simulated by BOMD. To induce chemical reactions, high initial collision energies between two compounds are added in BOMD. In this implementation, we introduced the prescreening step before BOMD, by using the virtual collision dynamics with atomic distance monitoring. This largely reduces the number of non-reactive trajectories in BOMD simulation and saves computational cost substantially in the treatment of reactions involving medium-sized compounds. After BOMD, we locate reaction events in trajectories using the HMM analysis of modified atomic connectivity matrix. Starting from several geometries close to reaction events, we try to build the elementary reaction pathway. In the TS searching step, different TS optimization manners are considered. We either directly take a few geometries nearby reaction events to conduct the TS optimization, or use the optimized intermediates (local minimum-energy geometries along reaction pathways) as the starting points in the TS searching. In the latter case, the climbing image NEB (CI-NEB)[64] approach is employed for the automatic TS location. Then, the vibrational analyses of TS geometries and intrinsic reaction coordinate (IRC)[65,66] calculations help us to identify proper TS structures and construct corresponding reaction pathways. Finally, the multi-reaction mechanism is constructed.

We employ the current theoretical approach to study two complicated reactions, *i.e.,* a penicillin G anion + a H$_2$O molecule and a penicillin G anion + a OH radical, which play important roles in environmental chemistry. The presence of antibiotics in wastewater is one of the crucial environmental problems and many researchers have paid considerable attention to their chemical conversions and removals.[67-74] Penicillin

G is one of the antibiotics widely used for human beings and animals in the treatment of various infections caused by bacteria. However, because of heavy use, it has been frequently detected in natural aqueous environment due to wastewater pollution, and thus its chemical reaction also received research interests.[75,76] Previous study mentioned that most penicillin G compounds in water exist as anion forms once pH value is higher than its $pK_a$ ($pK_a$ = 2.75).[76] This indicates that the study of the chemical reactions of penicillin G anions in aqueous environment is significant. Thus, we study two typical reactions, namely a penicillin G anion + a $H_2O$ molecule and a penicillin G anion + a OH radical. The importance of the first reaction is indubitable. The second reaction is also essential, as OH radical plays a critical role in photocatalysis decomposition of penicillin G in aqueous environment. As the additional tests of the proposed theoretical approach, we also investigated two simple reaction systems, $NH_3$ + $CH_3CHO$ reactions and $NH_3$ + HCHO reactions, which were studied by the work of Zimmerman.[26,77]

The current theoretical approach allows us to explore the multi-reaction mechanism of bimolecular systems. Our long-time goal is to provide a highly-automatic black-box approach to predict the unknown reaction products for given reaction species. As our first effort toward this goal, we show that the complicated bimolecular reaction mechanism may be identified by the current theoretical approach. In the current stage, even if we still cannot get the whole reaction network in a fully automatic manner, the proposed approach is still rather efficient. At the same time, this should be useful for researchers in chemical-relevant fields, improving the impact of theoretical chemistry on other research fields.

## 2. Theoretical Methods

A theoretical approach is developed for the automatic exploration of the bimolecular reaction mechanism, which combines various theoretical approaches, including on-the-fly reactive molecular dynamics, HMM, TS optimization and IRC.

Our purpose is to study bimolecular reactions and herein two reactant species are labelled as MOLE_A and MOLE_B in below illustration. Based on natural bond orbital (NBO) charges[78] and condensed Fukui function (CFF)[79] values, the reactive regions of reactant compounds are located. After that, the collision dynamics between two species is simulated using the on-the-fly BOMD. In this step, the dynamics is set to enforce the collision between the selected reactive regions of MOLE_A and MOLE_B. To find as many reactions as possible, different initial collision energies are taken in the BOMD simulation. For a balance of computational cost and accuracy, the BOMD simulation is run at a low-cost electronic-structure theory. To further reduce the computational cost of the dynamics simulations, we apply the preliminary virtual dynamics purely based on interatomic distances before the BOMD simulation. This allows us to get rid of the significant amount of non-reactive trajectories, and thus largely reduces the number of BOMD trajectories. After the BOMD simulation, the reaction events are identified by the HMM analysis of the atomic connectivity matrix. Starting from several geometries nearby the identified reaction time, intermediates are optimized first, which in fact correspond to the local minimum-energy structures along the reaction pathways. If two intermediates are connected by a barrier, they become the reactant and product of an elementary reaction. In this sense, there is no strict distinguishment between intermediates and reactants/products. Next, the TS search is performed using the standard TS optimization or the CI-NEB calculation. After the location of possible TS, the vibrational frequency analysis is conducted for validation. The IRC-driven pathways are built to connect reactants, TS and products.

All theoretical frameworks are discussed below and the computational details are given in the next section. The additional implementation details are also given in Supporting Information (SI).

**2.1 Reactive region selection**

When the medium-sized compounds are taken into account, the brute-force search of possible bimolecular reactions through collisions by BOMD may be extremely time consuming, because many trajectories are non-reactive. Thus, some tricks must be taken to avoid the non-reactive trajectories in the BOMD step, and at least we should make sure that the BOMD simulations are mainly run for the trajectories experiencing the effective collisions between reactive regions of MOLE_A and MOLE_B. For this purpose, it is quite crucial to identify the possible reactive regions of each compound.

Several descriptors can be employed to judge the reactive regions. The consideration of these descriptors allows us to select reactive regions automatically.

The NBO charge (reflecting the positive and negative net charge of each atom) and CFF (judging the electrophilic and nucleophilic attacking sites) values are used in the current approach. Here the condensed Fukui function (CFF) is employed to guide reaction region identification, which is the local representation of the original Fukui function.[80,81] More details on the selection of the reactive regions based on the NBO charges and CFF values are given in SI. Other descriptors may also be feasible choices. Certainly, it is also possible to apply the chemical intuition to choose reactive regions.

**2.2 Reactive dynamics simulations**

After the selection of reactive regions for MOLE_A and MOLE_B, we run the dynamics that gives the collision between them.

**2.2.1 Initial sampling**

In this step, the proper sampling of initial inter-molecular velocity is necessary to make sure that effective collisions appear in the dynamics simulation. Different initial collision energies are sampled to find relevant reactions as many as possible. More technical details on sampling inter-molecular motions and adding collision energies are given in SI.

For initial molecular orientation, we first fix the coordinates of MOLE_A. The collision dynamics should be aligned along with the direction connecting two individual centers of different reactive regions. All possible orientations of MOLE_B are sampled and the angular momentum is not considered. We wish to emphasize that molecular rotation generally takes place at a much longer time scale, for instance at picosecond time domain, while the current collision dynamics happens in the much shorter time scale. Thus, the sampling of angular momentum is not necessary. At the same time, we know that the orientation of molecule is extremely important for the reaction dynamics, thus we perform the sampling of different molecular orientations. In this way, the orientation-dependent reactions can be discovered. More details on how to perform the sampling of different molecular orientations are given in SI. Because the collision dynamics is used only for the generation of possible reactions, the current sampling approach should be acceptable.

The normal modes are obtained for both MOLE_A and MOLE_B individually. The Wigner sampling[82] of the lowest vibrational level is performed to create initial coordinates and velocities.

**2.2.2 Virtual dynamics**

In principle, it is possible to run the BOMD from all sampled initial conditions. However, this gives many non-reactive trajectories that require a huge amount of computational cost. Thus, additional efforts are taken to avoid them in the BOMD stimulation and here we introduce the virtual dynamics for prescreening purpose.

The virtual dynamics simulation is realized as follows. Starting from an initial sampling condition with only collision energy included, we run the virtual dynamics simulations without potential energy part, and check the inter-atomic distance at every timestep. If an atom in the reactive region of MOLE_A meets one atom in the reactive region of MOLE_B and no other atom is involved before their collision, we believe such trajectory is effective and should be re-run with the further BOMD simulation.

Otherwise, the trajectory is assigned as non-reactive trajectory that is not taken into account in the next BOMD simulation. The idea of the virtual dynamics shares some similarity with the discrete molecular dynamics (a kind of high-energy limit of BOMD)[83] widely used in biological system simulation. The computational setups are given in **Section 3** and the additional implementation details are given in SI**.**

**2.2.3 BOMD**

After initial sampling and virtual dynamics, the selected initial conditions are taken to run the on-the-fly BOMD simulations. Because different reactive regions, initial collision energies and initial orientations are considered here, the BOMD simulation step still requires a large number of trajectories. To save the computational cost, the BOMD should be run at a less computationally demanding electronic structure method. This setup is acceptable, because the BOMD step is only for the initial assessment of reactions. More additional discussions on the BOMD step are given in the next section and the SI.

**2.3 Reaction mechanism construction**

**2.3.1 Identification of the reaction events with modified connectivity matrix and HMM**

To find elementary reaction steps in BOMD trajectories, the bond cleavage and formation are detected based on the modified inter-atomic connectivity matrix and HMM.

The element of the atomic connectivity matrix $A$ is defined based on the inter-atomic distances as the following:

$$A_{ij} = \begin{cases} 1 & R_{ij} \leqslant 1.4 r_{ij} \\ 0 & otherwise \end{cases} \quad (1)$$

where $R_{ij}$ is the distance between two atoms, $r_{ij}$ is the standard covalent bond length of the two atoms, which is taken from reference.[84] Here the upper limit defines the threshold distance to divide the bound and unbound statuses, and the value 1.4 is taken from previous works.[53,54,63]

In the current BOMD simulation, the high initial collision energy is given, which indicates that the inter-atomic distance may experience a large amplitude motion even without reaction events. Thus, the employment of the connectivity matrix may provide many reaction events. If not filtered, a large number of structures should be taken into account in the next intermediate and TS optimization, largely reducing the efficiency of next steps. To remove the 'fake' reaction events in this step, the approach based on a two-state HMM is taken to analyze the atomic connectivity.

HMM is widely used in many research fields[53,54,63] and we only briefly outline the main idea here. Let us consider a set of discrete states $\{X_0, X_1, ..., X_t\}$ in a sequence of time $t$ that cannot be detected directly. This time series of "hidden" states generates a set of visible output states $\{V_0, V_1, ..., V_t\}$. A simple way to understand the two-state Markov chain model is shown in Figure 1 (a). For a two-state HMM, the initial probability vector is $[P(\text{State 1}), P(\text{State 2})]$. The time evolution of the probability vector is given by the transition matrix

$$T = \begin{pmatrix} 1-\alpha & \alpha \\ \beta & 1-\beta \end{pmatrix} \qquad (2)$$

where $T = P(X_t|X_{t-1})$ and $1-\alpha$ ($1-\beta$) defines the probability of transition between different states. At each time step, the probability of output state observed is given by

$$O = \begin{pmatrix} 1-\lambda & \lambda \\ \mu & 1-\mu \end{pmatrix} \qquad (3)$$

where $O = P(V_t|X_t)$, $\lambda$ ($\mu$) represents the probability to assign State 1 (State 2) as its own, while $1-\lambda$ represents the probability of the wrong assignment. Figure 1 (b) shows the relationship between 'hidden' states, observation states, transition probability and output probability.

In the reaction detection step, the connectivity matrix is calculated to identify whether a bond is broken or not. In principle, the hidden states can be viewed as the identification on whether the bond status changes or not for any pair of atoms. However, many chemical bond statues may vary at the same time and thus it is necessary to monitor the overall result. In implementation, when no bond breaks or forms, we assume that the hidden state does not change. Otherwise, the hidden state changes. The observable states now are whether the connectivity matrix changes or not. More additional discussions on the HMM model are given in SI.

The initial probabilities of two hidden states can be set as 0.5. The transition matrix is given by

$$T = \begin{pmatrix} 0.9999 & 0.0001 \\ 0.0001 & 0.9999 \end{pmatrix} \qquad (4)$$

and the output probability matrix is

$$O = \begin{pmatrix} 0.55 & 0.45 \\ 0.45 & 0.55 \end{pmatrix} \quad (5)$$

Based on the HMM analysis of the atomic connectivity matrix, the reaction events in trajectories are identified. The values in the transition matrix $T$ and output matrix $O$ are not the same as the ones used in previous works that ran reactive dynamics at high temperature.[53,54,63] The current work only puts the high kinetic energy in the direction of collision between two reactive regions initially and the inner energy of each molecule is still low. In this way, the molecular geometry is not far from the ground-state minima and no strange structures appear before collision. However, the excessive collision energy induces the large-amplitude bond stretching motions during collisions. In this situation, two originally unbound atoms may quickly become very close to each other, and then return to their unbound status again immediately. Such features appear quite often in the collision dynamics, and it is necessary to get rid of them in the identification of reaction events. For this purpose, a modified connectivity matrix is proposed as

$$B_{ij} = \begin{cases} 1 & \frac{r_{ij}}{1.4} \leqslant R_{ij} \leqslant 1.4 r_{ij} \\ 0 & otherwise \end{cases} \quad (6)$$

The elements of the $B$ matrix, $B_{ij}$, displays the ultrafast oscillation patterns, when such high-energy non-reactive collision happens. If we can detect such high-oscillation feature in the HMM analyses, such 'fake' events can be filtered out. For this purpose, we use the $B$ matrix instead of the original $A$ matrix to conduct the HMM analyses. Also for the same purpose, we adjust the values of two matrices ($T$ and $O$) in HMM. This explains why the parameters used in the $T$ and $O$ matrices are different to previous works.[53,54,63] At the end, the current setup makes sure that only one reaction event is detected for most trajectories.

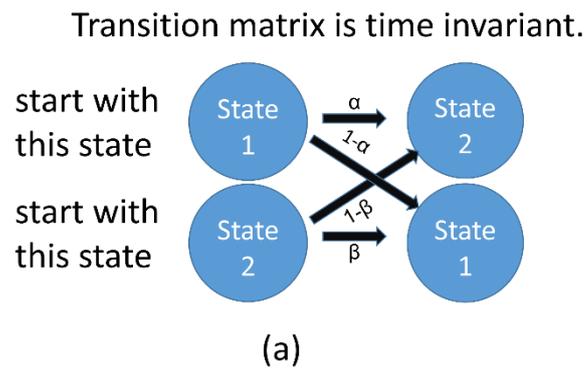

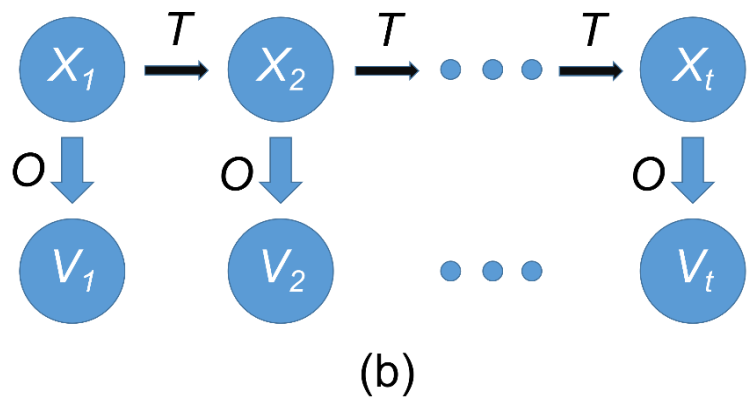

Figure 1. (a) is the two-state Markov model with transition probabilities $\alpha$ and $\beta$; (b) gives the transition progress of Chain $X$ and the output progress of visible sequence $V$. $T$ is the transition matrix and $O$ is the output matrix.

**2.3.2 Transition state search**

When the reaction time $t_r$ is determined, a time interval [$t_s$, $t_f$] centered at $t_r$ is defined. Then a few of geometries equally distributed in this time interval are chosen as starting guesses to perform the intermediate optimization. In this situation, two time-adjacent selected geometries may give the same or different optimization results. In principle, an elementary step should take place in the latter case. After optimization, the latter case can be identified by checking the atomic connectivity matrix and this sub-time interval (labelled as [$t_i$, $t_{i+1}$]) containing the elementary reaction step is known. Next, several geometries in the trajectory propagation within this sub-time interval are taken to perform the TS optimization. Alternatively, the intermediates optimized from the geometries at $t_i$ and $t_{i+1}$ are chosen to perform the CI-NEB task.

After the TS optimization, the atomic connectivity matrix is used here to divide all optimized structures into several groups. To avoid repeated calculations, the frequency analysis is conducted for the representative structures of each obtained TS groups. The real TS structures are identified, when only one imaginary frequency is observed and the vibrational motion of this mode is localized in the reactive region.

**2.3.3 IRC and reaction mechanism construction**

IRC calculations are performed for all real TS geometries. The end points of the IRC path are optimized to compare with the reactant/product for an elementary reaction.

This step makes sure that the obtained reaction pathways are dynamically accessible. At the end, the connection of all reaction species gives the multi-reaction mechanism.

## 3. Computational Details

We mainly considered the bimolecular reactions of two systems, namely a penicillin G anion + a $H_2O$ molecule and a penicillin G anion + a OH radical reactions. As a complimentary study, we also explored two simple models, $NH_3$ + $CH_3CHO$ reactions and $NH_3$ + HCHO reactions, which was studied in previous works.[26,77]

Initially, all electronic structure calculations were performed using the Gaussian 16[85,86] package. The geometry optimization, frequency calculations and NBO charge analysis were performed at the B3LYP[87,88]/6-31G(d,p)[89-100] level. The CFF values were calculated by Multiwfn[101] software based on Gaussian output. More technical details on how to perform the selection of reactive regions are given in SI.

In the virtual dynamics, we originally sampled 20000 initial conditions with different initial orientations for each pair of reactive regions. The additional discussion on the orientation sampling in the virtual dynamics is given in SI. The initial distance between them was set as 7 Å. For each pair of reactive regions, about 5000 initial conditions may give effective collisions.

After the virtual dynamics simulation, 700 initial conditions randomly chosen from the above effective samples were taken in the BOMD simulations. In BOMD run, the nuclear motion was integrated with the velocity-Verlet algorithm. The BOMD simulation was run up to 100 fs with the time step of 0.5 fs. The BOMD dynamics was run at the HF level with the 3-21G[102-107] basis set using our home-made code interfaced with Gaussian 16 package.

To avoid the missing of low-energy reactions, different collision energies were chosen in the BOMD simulation. The threshold between low-energy and high-energy reactions was set to be a rather high value 40 kcal/mol for reaction barrier. The reaction with such high barrier is very difficult to take place at the room temperature. In this implementation, we firstly ran the preliminary BOMD with one collision energy to check the reaction ratio. Next, we adjusted the initial collision energies to re-run the BOMD. Here, we first ran some test trajectories (~20) with some initial energies. If no reaction happened, we just enlarged the collision energy by a factor of 1.2 to run the test collision dynamics again. The collision energy range in these test calculations was 0.37-0.66 Hartree. For each selected collision energy, we firstly ran 500 trajectories to obtain the main results and another 200 trajectories were calculated to check convergence. When no new low-energy reaction was found in these additional 200 BOMD trajectories, we believed the calculation results are converged at this collision energy. Next, we needed to examine the low-energy reactions under different initial collision energies. Once no new low-energy reaction was found, the calculation was thought to be converged and we did not increase the collision energies any more.

The modified atomic connectivity matrix was calculated along the trajectory propagation. The HMM was employed to select the reaction events. At each reaction time, we defined a time interval ~30 fs, and chose six geometries equally distributed within such interval for the intermediate optimization, see $S_1$-$S_6$ in Figure 2. If two time-adjacent structures (for instance $S_i$ and $S_{i+1}$) give different optimization results, we may assume that this sub-time interval contains an elementary reaction step. In the current work, we fixed the time interval to 30 fs. Although the optimization of this value is not considered in the current work, this value is acceptable because we got enough number of the TS structures (see below). Certainly, the dependence of the results on this time interval may be an interesting topic in the future research.

Next, three geometries were chosen within the time interval from $S_i$ to $S_{i+1}$ to perform the direct TS optimization at the B3LYP/6-31G(d,p) level using Gaussian 16. If the TS optimization was not converged, we tried to employ the CI-NEB approach to redo the TS search again, in which two intermediate geometries optimized from $S_i$ and $S_{i+1}$ were taken as the starting and ending points. The CI-NEB work was conducted at the B3LYP/6-31G(d,p) level using Orca.[108] Here eight images were used along the guess reaction pathways and all other details in the CI-NEB calculations followed the standard setups in Orca.

After the collection of all optimized TS geometries, the frequency analyses and IRC calculations were performed to build the reaction mechanism, which were all conducted at the same level of theory using Gaussian 16. To make sure that the IRC results correctly reflect the reactions given by the BOMD, we also examined the geometry similarity between the optimized endpoints of IRC and two intermediate geometries optimized from $S_i$ and $S_{i+1}$, by comparing their connectivity matrices. At the end, all species were connected to build the reaction network.

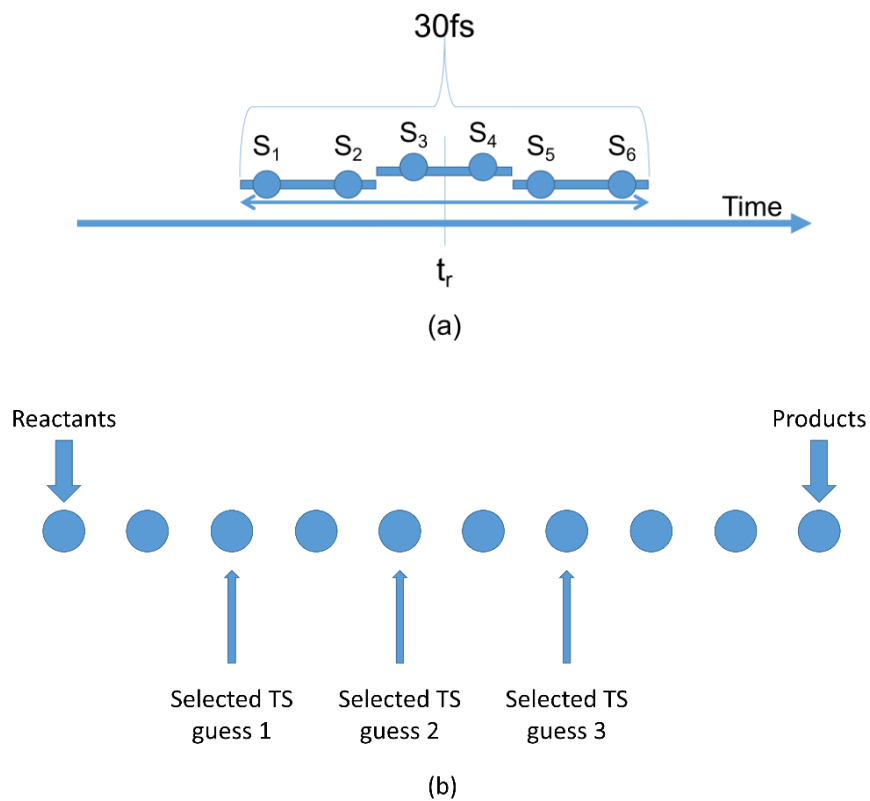

Figure 2. In the 30-fs time interval around the reaction event at time point $t_r$, six equally distributed structures (labeled as $S_1$-$S_6$ in (a)) were taken as the initial guesses to perform the intermediate optimization. If reactants and products of one elementary step were obtained from optimization jobs, three geometries within the sub-interval were chosen for the TS optimization.

## 4. Results

Here we examine the feasibility and applicability of our proposed theoretical approach by studying two typical penicillin G anion-involved reactions that are environmentally significant. The first one is penicillin G anion + $H_2O$ reaction. And the second one is the penicillin G anion + OH radical reaction. For illustration, we label penicillin G anion as MOLE_A and the second specie ($H_2O$ or OH radical) as MOLE_B. As a complimentary study, we also tested the performance of the current approach by using another two simple models, $NH_3$ + $CH_3CHO$ reactions and $NH_3$ + HCHO reactions, which were studied by the work of Zimmerman.[26,77] Since these two systems are rather small and only used for additional test, we put their results in the SI and mainly focus on two medium-sized compound reactions. All energies shown in this work are given in the electronic energies, instead of Gibbs energies. This gives the rather qualitative understanding of the reaction mechanism.

**4.1 Reactions between penicillin G anion and $H_2O$ molecule**

The optimized geometry of penicillin G anion is given in SI Figure S1. The NBO charges and CFF values are shown in SI Table S1. As shown in Figure 3, three reactive regions are selected for MOLE_A, as the atoms in these regions may be involved in reactions due to large NBO charges and CFF values. Reactive region 1 contains the peptide bond. Reactive region 2 contains the amide bond of the lactam moiety. Reactive region 3 contains the $COO^-$ group. The whole $H_2O$ molecule is set as the reactive region of MOLE_B, since both O and H atoms are active.

In principle, it is also possible to include more atoms into the reactive regions. However, the current calculations show that all obtained elementary reactions between the reactive regions of MOLE_A and H₂O already display rather high energy barriers. At the same time, the reactivity of other non-selected atoms is much weaker according to their NBO charges and CFF values. This indicates that the reaction probability between the non-selected atoms and H₂O should be extremely small. Thus we believe the above selection should already cover the primary reactions.

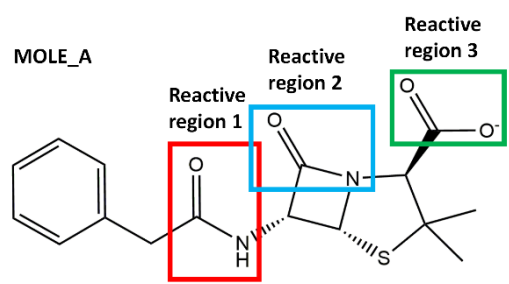

Figure 3. Three reactive regions are selected for MOLE_A. Reactive region 1 contains the peptide bond (red zone); Reactive region 2 contains the amide bond of the lactam moiety of the four-membered ring (blue zone); Reactive region 3 contains the COO⁻ group (green zone).

Following the automatic reaction-mechanism discovery procedure described above, totally nine elementary reactions are obtained, as shown in Figure 4.

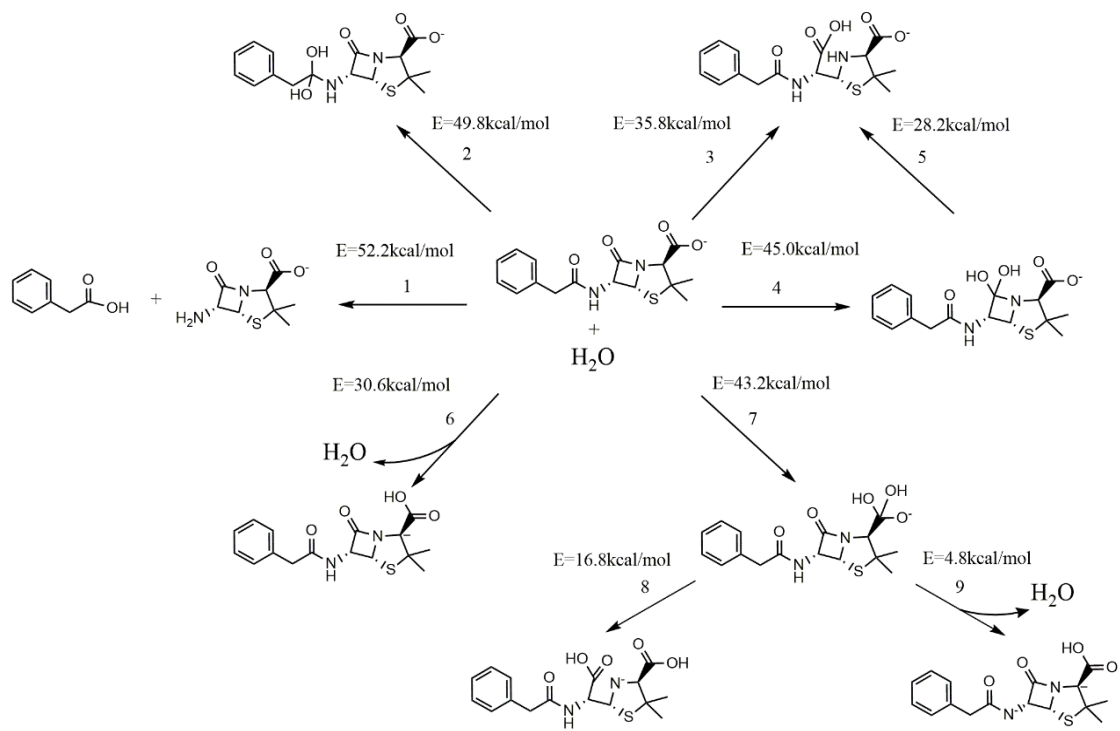

Figure 4. Network of penicillin G anion and H$_2$O reactions. The reaction barrier height (with the unit of kcal/mol) is given for each reaction. All elementary reactions are labeled by a number here in order to refer them as Pathway 1-9 in the mechanism discussion.

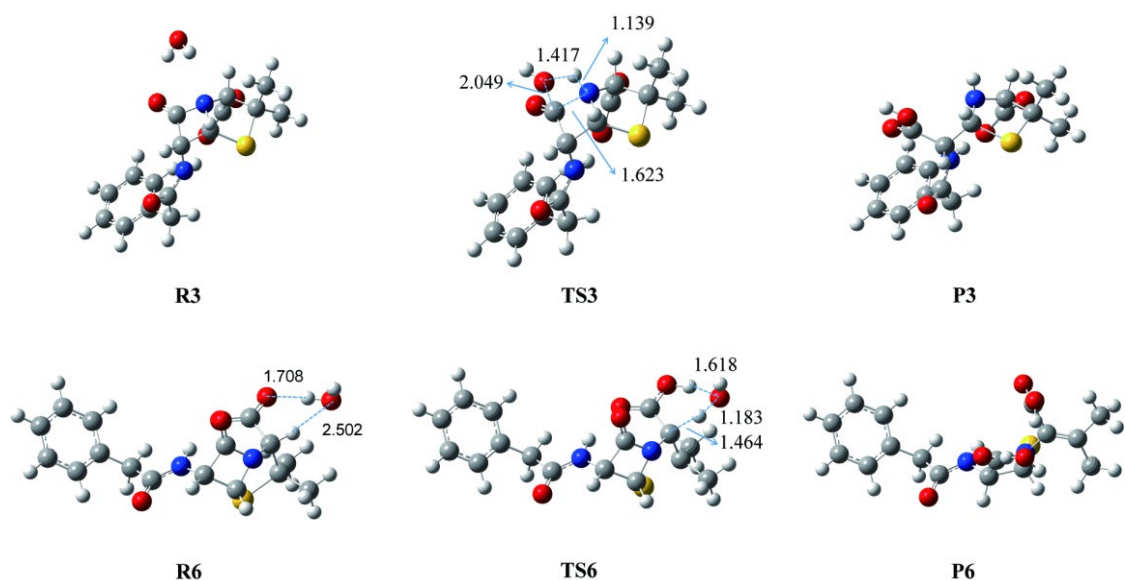

Figure 5. Important species involved in the penicillin G anion + H$_2$O reactions shown in Figure 4. Other involved species are shown in SI. Here we label each structure

according to the following rule: the capital letters, **R**, **TS** and **P** that denote reactants, TS, and products respectively. The numbers after these capital letters correspond to the reaction indices given in Figure 4. The dashed line and the blue arrows show the key atomic distances with the unit of Å.

Two types of reactions are found, which are the hydrolysis reactions (Pathway 1-5, 7 and 8 in Figure 4), and the H-atom exchange reactions (Pathway 6, 7 and 9 in Figure 4). Here Pathway 7 is involved as the first step in both the hydrolysis (Pathway 7 to Pathway 8) and H-atom exchange (Pathway 7 to Pathway 9) reactions.

For hydrolysis reactions, the reaction site involves either the peptide bond in Reactive region 1 of MOLE_A or the amide bond of the lactam moiety in Reactive region 2 (Figure 3). Here the $H_2O$ molecule attacks the C-N bond in reactive regions, As the result, both the C-N and O-H bonds break, while the C-O and the N-H bonds are formed in products. The above bond cleavage and formation may take place at the same time, giving two reactions (Pathway 1 and Pathway 3 in Figure 4). Alternatively, it is also possible that the O atom of $H_2O$ connects to the C atom of the C-N bond firstly, and then the further reaction takes place, resulting in the two-step reactions (Pathway 4 to Pathway 5, and Pathway 7 to Pathway 8 in Figure 4).

Among all hydrolysis reactions, the one-step hydrolysis reaction of the lactam region (Pathway 3 in Figure 4) shows the lowest reaction energy barrier of 35.8 kcal/mol, and thus we take this reaction as a typical example for illustration. All critical geometries (**R3**, **TS3** and **P3**) in this elementary reaction step are given in Figure 5. Firstly, one H atom of $H_2O$ moves towards the N atom of the amide bond in the lactam region and the O-H bond starts breaking. Then the other OH group of $H_2O$ accesses to the C atom of the amide bond and induces the C-N bond cleavage. At the TS (**TS3** in Figure 5), the elongations of the H-OH (1.417 Å) and C-N (1.623 Å) bonds are observed, while both the N-H and C-O bond distances become shorter, which are 1.139 Å and 2.049 Å, respectively. At the end, two bonds (H-OH and C-N) break and two new bonds (C-O and N-H) are formed, and the product is given as **P3** in Figure 5. Other one-step and all two-step hydrolysis reactions display rather high barriers, so we do not discuss them here.

For H-atom exchange reactions, two reaction mechanisms are found. The first one denotes Pathway 7 to Pathway 9 in Figure 4 (Reactive region 1) and the second one is Pathway 6 in Figure 4 (Reactive region 3). The latter one displays the lower barrier (30.8 kcal/mol), and thus the detailed mechanism is given here. In this reaction, the $COO^-$ group takes the H atom of $H_2O$, and the remaining OH part of $H_2O$ captures the H atom originally connecting to the α-C atom in adjacent to the $COO^-$ group. At the TS structure (**TS6** in Figure 5), the relevant O-H and C-H bond lengths become longer, 1.618 Å and 1.464 Å. At the end, the H-atom exchanging reaction is finished and the $H_2O$ molecule is re-generated, see products **P6** in Figure 5.

All penicillin G anion + $H_2O$ reactions display rather high energy barriers, indicating that these reactions are not easy to take place at the room temperature. Given the high stability of the penicillin G anion in aqueous environments, we may explain why such compounds can be detected in wastewater and become a group of typical pollutants. Thus, some additional efforts must be taken for their management, for instance by adding some chemicals or using photocatalysis reactions.

**4.2 Reactions between penicillin G anion and OH radical**

In the second example, we study the reactions between penicillin G anion and OH radical. Because of the high reactivity, the OH radical should easily react with more regions of penicillin G anion. Therefore, we only use CFF values here to choose the reactive regions of the MOLE_A. As shown in Figure 6, five reactive regions are selected for MOLE_A. Reactive region 1 includes the six-member aromatic benzene ring. Reactive region 2 contains the peptide bond. Reactive region 3 contains the amide bond of the lactam moiety of the four-membered ring. Reactive region 4 contains the COO$^-$ group and Reactive region 5 refers to the C-S moiety in the five-member ring. The whole OH radical is selected as reactive region for MOLE_B.

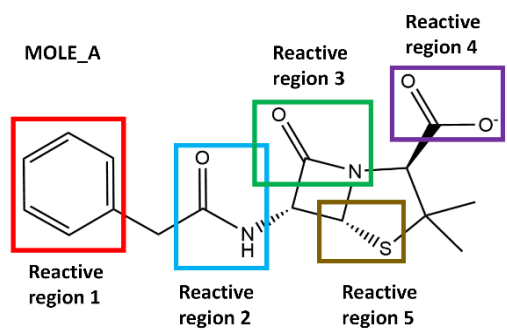

Figure 6. Five reactive regions are selected for MOLE_A: Reactive region 1 includes the six-member aromatic benzene ring (red zone); Reactive region 2 contains the peptide bond (blue zone); Reactive region 3 contains the amide bond of the lactam moiety of the four-membered ring (green zone); Reactive region 4 contains the COO$^-$ group (violet zone) and Reactive region 5 refers to the C-S moiety in the five-member ring (brown zone).

Totally nineteen different reaction pathways are found for penicillin G anion + OH radical reactions, as shown in Figure 7. These reaction pathways are roughly divided into four types. In Type-I reactions, the OH radical simply attaches to one atom of MOLE_A. In Type-II reactions, after the OH radical connects to one atom of MOLE_A, the additional bond cleavage or formation happens, or the second-step reaction occurs. In Type-III reactions, the OH radical takes one H atom away from penicillin G anion and H$_2$O is formed. Type-IV reactions lead to the bond cleavage of OH radical. Several critical geometries are shown in Figures 8.

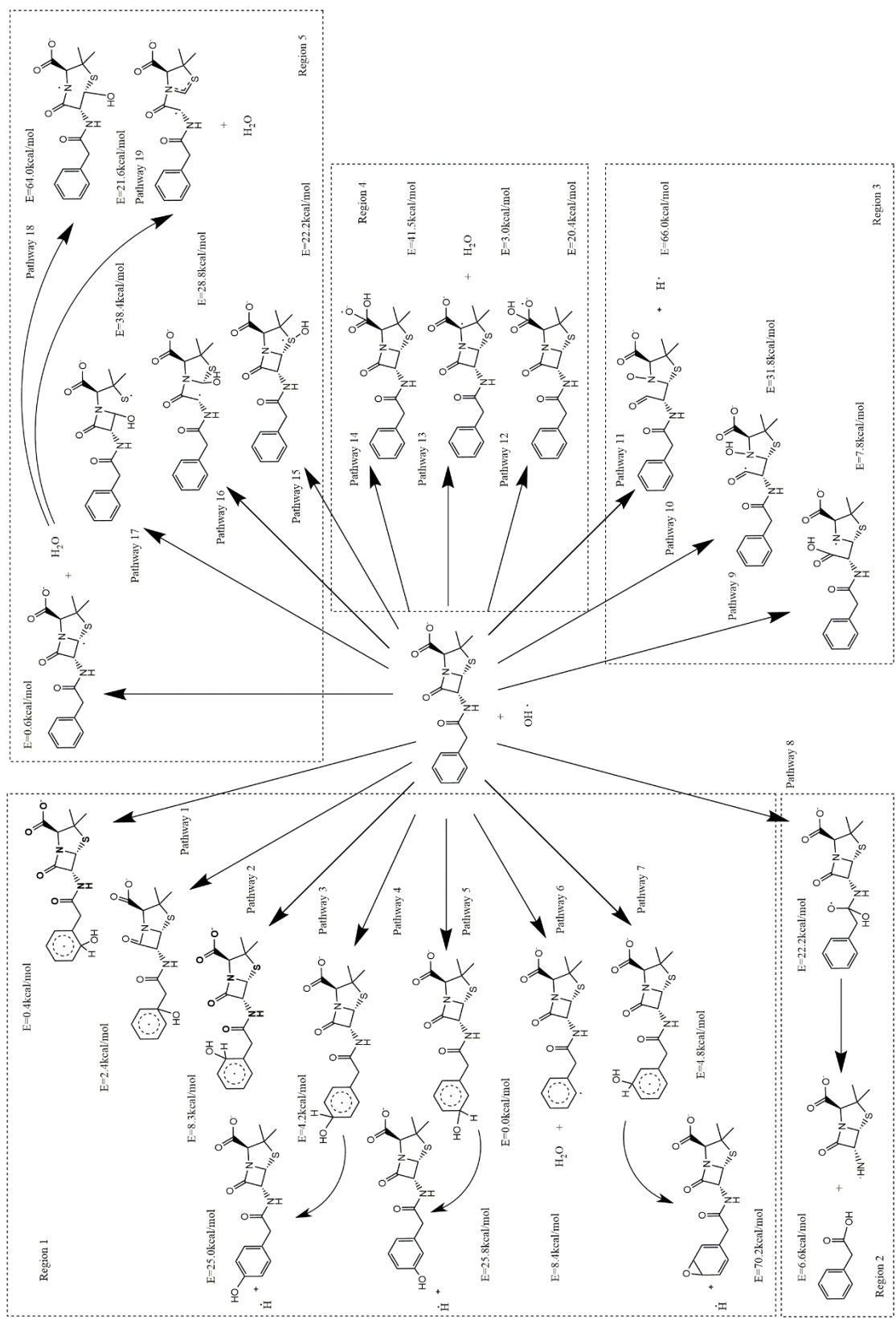

Figure 7. All penicillin G anion + OH radical reactions. Each reaction pathway is labeled by its index number (*i.e.,* labeled as Pathway 1 to Pathway 19) in illustration

and reaction barrier height is given in the unit of kcal/mol.

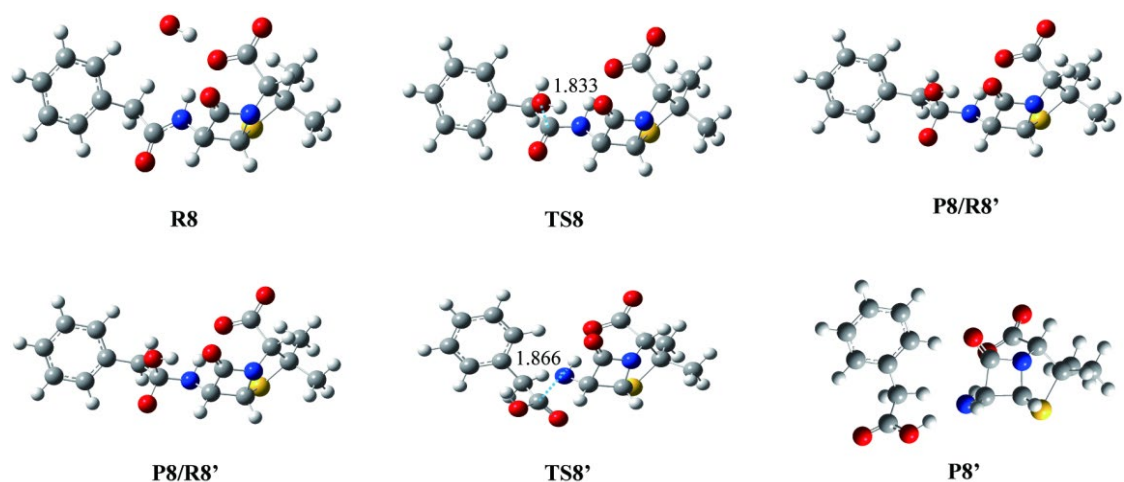

Figure 8. Several critical geometries in the penicillin G anion + OH reactions shown in Figure 7. The dashed line and the blue arrows show the key atomic distances with the unit of Å. Other involved species are given in SI.

All four types of reactions are found for the benzene moiety (Reactive region 1). As shown in Figures 6-8, Pathways 1, 2, and 3 are all Type-I reactions with low-energy barriers, in which the OH radical simply connects to one C atom in the six-membered ring. Pathway 1 may be the barrierless reaction with $E$ ~0.4 kcal/mol. As Type-II reactions, both Pathways 4 and 5 are composed of two steps. The OH radical connects to one C atom in the ring, and then the H atom originally connecting with this C atom flies away. The only Type-III reaction detected in this region follows Pathway 6, in which the OH radical captures the H atom of six-member ring and the $H_2O$ molecule is formed. The Type-IV reaction here follows Pathway 7, which includes two elementary steps. After the OH radical connecting with the C atom in the ring, the H atom of this OH radical simply goes away, while the very high barrier ($E$ ~70.2 kcal/mol) seems to prevent this pathway.

Only one two-step reaction exists in the peptide bond region (Reactive region 2), which is assigned as the Type-II reaction (Pathway 8). The first-step reaction denotes that the OH radical connects with the C atom in the peptide bond. In the second step, the C-N bond becomes longer (1.866 Å at **TS8'** in Figure 8) and the C-N bond breaks (**P8'** in Figure 8). The first step is the rate-deterministic step with the reaction barrier of 22.2 kcal/mol. Although this reaction may be accessible, it should play a minor role, because many lower-energy reaction pathways exist in other reactive regions.

Three reaction pathways are presented in Reactive region 3 and both C, N atoms in the amide bond of the lactam moiety become reaction sites. Among them, Pathways 9 and 10 belong to Type-II reactions and Pathway 11 belongs to Type-IV. The OH radical connects to either the N atom (in Pathways 10 and 11) or the C atom (in Pathway 9) of the C-N bond, and then the C-N bond breaks. The different between Pathway 10 and Pathway 11 is that the original OH bond breaks in the latter case. Among all reaction pathways in this region, Pathway 9 shows the lowest reaction barrier with 7.8 kcal/mol.

Three reactions take place in Reactive region 4. The Type-I reaction is governed by Pathway 12 in Figure 7, in which the OH radical directly connects to the C atom in $COO^-$ group. Pathway 13 gives the Type-III reaction, in which the OH radical attacks to the H atom connecting to the α-C atom in adjacent to the $COO^-$ group and takes this H atom away to form H$_2$O molecule. Pathway 14 belongs to the Type-IV reaction, in which the OH radical moves close to one C=O bond in the COO$^-$ group firstly. After the O-H bond breaks, the O atom connects to the C atom of the COO$^-$ group and the H atom connects to one O atom of the COO$^-$ group. In this region, Pathway 13 shows the lowest energy barrier of 3.0 kcal/mol.

In Reactive region 5, five reactions are found. Pathways 15 and 16 are assigned as the Type-I reaction, in which the OH radical simply connects to the C atom and the S atom of the C-S bond, respectively. Pathway 17 (Type-II) gives a high-energy barrier reaction, in which the OH radical connects with the C atom of the C-S bond, and this results in the C-S bond breaking. Both Pathways 18 and 19 (Type-III) display the same first reaction step, *i.e.,* that the OH radical captures the H atom connecting to the C-S bond and the H$_2$O molecule is formed, while different reaction mechanisms are found in their second reaction steps. In Pathway 18 the C-N bond in the lactam moiety breaks, while in Pathway 19 only the C-C bond of the four-member ring breaks.

Overall, several low-energy reaction pathways are found in the penicillin G anion + OH radical reactions, due to the high reactivity of the OH radical. Thus, it is very easy to induce the chemical transformation of the penicillin G anion under the existence of OH radicals.

## 5. Discussion

Here we discuss several aspects of the current theoretical approach for the automatic exploration of reaction mechanism.

One great advantage of this theoretical method is that it can naturally be implemented in the parallel way, because different trajectories and TS finding tasks can be run independently. The price is that more convergence tests are necessary for the determination of some simulation setups, such as selecting reactive regions, choosing initial collision energies and so on.

The reactive-region selection and the virtual-dynamics step are very essential, because they allow us to run much smaller number of trajectories in the BOMD step. These tricks largely reduce the number of non-reactive trajectories in the BOMD step and save a huge amount of simulation time. When the reaction involves the large-sized compounds, such approaches become extremely useful to reduce computational cost. Certainly, great caution should be paid to avoid the loss of important reactions. Thus, the preliminary test calculations on how to set up these prescreening calculations are helpful.

In the selection of the reactive regions, the NBO charges and CFF values are employed and more details are given in the SI. We also noticed that the previous work by Reiher and co-workers also chose to use the similar molecular descriptors to define the reactive sites.[29] Certainly, the current selection standard may not always be optimal. For instance, we may directly use some loose thresholds to include more atoms or try different criteria. These setups can always be examined by checking whether the observed low-energy reactions do not change with such parameters. At the same time, the involvement of chemical intuition may be highly helpful.

In a few situations, some reactions involve some atoms out of the reactive region, for instance, Pathway 6, 8 and 9 in the penicillin G anion + $H_2O$ reactions and Pathway 13 in the penicillin G anion + OH reactions. Such results may be attributed by a few reasons. For instance, the virtual dynamics and the real BOMD were run with different setups. In the BOMD dynamics the potential energy was considered and the vibrational level was sampled initially. In addition, for a two-step reaction, the second step may be fully determined by the product of the first step. Due to these reasons, a few reactions found by the current approach may involve some atoms out of the pre-defined reactive regions. However, such situations happen very occasionally.

In this work, the HF/3-21G level is used in BOMD part to generate the initial guesses for the next TS searching tasks. In fact, many BOMD-based approaches[46,53] used the low-cost electronic-structure methods, such as semiempirical method or HF, to speed up the trajectory propagation. This method in principle works here, because this step is only used to generate the initial guess structure for the successive reaction pathway construction. As a complimentary test of the employment of the HF/3-21G level in BOMD, we also studied the bimolecular reactions of small systems shown in the works by Zimmerman,[26,77] see SI for details. Our results on the low-energy reactions are more-or-less consistent with his previous findings, although some differences exist in the high-energy reactions. This indicates that our selection of the HF/3-21G level in the BOMD is reasonable. Certainly, the limitation of the employment of the HF/3-21G level is also clear, including that it may give the improper description on the reaction barriers, provide the inaccurate description on the bond dissociation limit and miss some important reactions.

Next, we employ the HMM analysis of the modified atomic connectivity matrix in the determination of the reaction events, avoiding the selection of a large number of non-reactive 'fake' reactions caused by high collision energies. This provides more reasonable initial guess for the intermediate and TS searching, and significantly improve the efficiency of the reaction pathway construction.

One BOMD trajectory can only follow a single pathway and does not give us two different reaction pathways. In principle, it can experience the reactions involving successive steps, while the connection between the multi-step reaction and BOMD trajectory is not trivial. Taking a two-step reaction as an example, it is certainly true that a single BOMD trajectory may experience both two steps. However, when we used the HMM as a filter, it is possible that only one reactive event is identified. Since we took a time interval centered at the located reactive event and chose several initial configurations for the successive intermediate and TS search, these geometries may already cover two reaction steps. As the consequence, we may find two successive reaction steps. In other situations, maybe only the first step or the second step was obtained from the reaction pathway construction. However, at the end, we need to connect all involved species to build a reaction network. In this step, we compared the similarity of all obtained species based on the inter-atomic connectivity matrix. When two species show the similar connectivity matrix, we assign them as the same intermediate. Thus, the same intermediate may be composed of different isomers for such large system. In the current work, we only put them together and did not pay attention on the relation between these isomers. Such limitation will be the research topic of our further work.

In the current work, the selection of collision energies is fully determined by the ratio of reactive trajectories in all trajectories. That's why we recommended to take the iterative procedure to determine the suitable collision energies. When no more low-energy reaction (energy barrier < 40 kcal/mol) is found with the increasing the collision energy, we believe that the calculations are converged. Normally different reaction products may be found under different collision energies. For example, with the increasing of the collision energy, we found that more trajectories gave high-energy reactions. For different reactive regions, different optimal collision energies may be chosen according to the tested results. The current iterative procedure normally gives the rather large collision energy ( > 0.37 Hartree). We also noticed that the very high kinetic energies were employed in other works of the BOMD-based reaction mechanism discovery.[46,54] For instance, the ab initio nanoreactor used ~ 2000K to run BOMD and the instantaneous temperature imparted by the piston reached to ~ 10000K.[54] And in the TSSCDS[46] method, the more than 20 kcal/mol (per normal mode) energy was initialized to run the BOMD. They obtained reactions with barrier height ~10-100 kcal/mol. In fact, it may be very difficult to give the direct connection between reactive barrier height and such high kinetic energies energy. The reason may be that the vibrational energy redistribution is generally extremely fast in the classical dynamics simulations. Even if we add the very high initial collision energies, such energy is quickly dissipated into the whole system due to the large molecular size. Thus, it is not easy to characterize the energy flow during the collision. The characterization of such fast energy flux is beyond the scope of the current work. We wish to emphasize that the reasonable reactive network is obtained with such collision energies. Thus, for practice purpose we believe such selection of the collision energy is reasonable.

We try to address the convergence of our calculations by examining whether the low-energy reactions can be found as many as possible in the current approach. In practice, with the increase of the collision energy, we found more trajectories gave high-energy reactions. It is also important to known whether there are enough trajectories used in the BOMD step. In practice, the iterative procedure was employed to know how many trajectories are required. For instance, we ran 500 trajectories first and then 200 trajectories additionally to check whether new reactions were found. This makes sure that the results are converged.

Table 1. Number of TS Structures *v.s.* Number of Trajectories (a Penicillin G anion + a OH radical, Reactive region 1)

| Collision Energy (Hartree) | Number of Trajectories | TS Numbers (different connectivity) | Reactions |
|---|---|---|---|
| 0.45 | 500 | 6 | 1,2,3,5,6 |
| 0.54 | 500 | 10 | 1,2,3,4,5,6,7 |
| 0.64 | 500 | 10 | 1,2,3,4,5,6,7 |

Taking the reactions between a penicillin G anion + a OH radical as example, we try to analyze all reactions in Reactive region 1, see Table 1. With the low collision energy (0.45 Hartree), totally 5 different reaction pathways were found for Reactive region 1, and all of them are low-energy reactions, including Pathways 1, 2, 3, 5 and 6 in Figure 7. With the medium collision energy (0.54 Hartree), all above reactions (Pathways 1, 2, 3, 5 and 6) appeared. And we found another two low-energy reactions, Pathways 4 and 7 in Figure 8. Under the high-level collision energy (0.64 Hartree), we did not find any additional low-energy reactions. In this sense, we think that most low-energy reactions should have been found and the employment of higher collision energy is not needed.

In principle the searching of possible reactions in the high-dimensional potential surface of a complicated system is a kind of so-called "NP-hard" problem that is basically computationally unsolvable. We can never make sure that 100% of low-energy reaction pathways can be found, no matter which theoretical approach is employed. All approaches, including the current one, only try to find relevant reactions as many as possible. For instance, in the test example of $NH_3$ + $CH_3CHO$ reaction provided by the previous work by Zimmerman,[77] we found the same two low-energy reactions, as well as several high-energy reactions, see SI for details. For the second example $NH_3$ + HCHO reaction, we also found the same lowest-energy reaction (32 kcal/mol). For the second lowest-energy reactions, although we also clarified it as a $H_2$ elimination reaction, the reactive barrier and products are different: 63 kcal/mol in this work and 80 kcal/mol in the work by Zimmerman.[26] In addition, the different products are found here. The products are $H_2$, CO and $NH_3$ in the work by Zimmerman, while they are $H_2$ and $NH_2CO$ in the current work. Thus, the low-energy reactions (< 40 kcal/mol) are more-or-less addressed by our approaches. When much larger systems are involved, such as a penicillin G anion + a $H_2O$ molecule and a penicillin G anion + a OH radical, the current approach finds many low-energy reactions in a rather automatic way, therefore we believe this protocol is useful.

In the current work, we only use the B3LYP/6-31G(d,p) level to optimize the TS geometries and build the reaction pathways. In principle, this electronic level is enough to get qualitative answers for our chosen reactions. If quantitative results are necessary, more accurate electronic-structure methods, such as DFT with more accurate functionals, the second-order Møller-Plesset and even couple-cluster approaches, become suitable choices in the TS searching. This may improve the accuracy in the description of each reaction mechanism, by giving more precise energies for all species. It is also possible to describe solvent effects by using the continuum solvation model.[105] All these additional improvements can be introduced, because any electronic structure method can in principle be applied in our current approach.

## 6. Conclusion

A theoretical methodology is introduced for the automatic discovery of bimolecular reaction mechanism. The current protocol employs several steps, including the reactive regions selection, the preliminary virtual dynamics, the successive BOMD simulation with different initial collision energies, the identification of reaction events with the HMM analysis of modified atomic connectivity matrix, the intermediate and TS searches, and the IRC construction.

First of all, reactive regions are selected for all reactants based on NBO charges and CFF values. The trial dynamics simulation, namely virtual dynamics with atomic distance monitoring, is used for the prescreening of reactive trajectories, and this largely reduces the number of non-reactive trajectories in the next BOMD step. After the BOMD simulations with different collision energies, the HMM analysis of the modified atomic connectivity matrix gives the rough estimation of the reaction events in each trajectory. A few of structures in adjacent of the reaction events are chosen for the optimization of all intermediates and TS. Both direct TS optimization and CI-NEB calculation are conducted here. When all TS structures are obtained, the frequency analyses and IRC calculations are employed to pick up the proper TS geometries relevant to true reactions. After the connection of all species, the whole reaction network is established.

The feasibility of the current theoretical approach is examined by two typical reactions (a penicillin G anion + a $H_2O$ molecule and a penicillin G anion + a OH radical), which play important roles in environment science. For the first example, penicillin G anion + $H_2O$ reactions, nine elementary reactions are identified, including hydrolysis and H-atom exchanging pathways. For the second example, penicillin G anion + OH radical reactions, a very complicated reaction network with nineteen reaction pathways is built. In the latter case, the complicity of the multi-reaction network is due to the high reactivity of the OH radical.

The current theoretical method shows both effectiveness and applicability in the computational study of bimolecular reaction mechanism involving intermediate-sized molecule systems. We believe that with the further development the automatic method may be a useful protocol in the exploration of other general chemical reactions in the future.

## ASSOCIATED CONTENT

**Supporting Information**

This information is available free of charge via the Internet at http://pubs.acs.org

(1) Optimized structure of penicillin G anion, NBO charges and CFF values of penicillin G anion, selection of reactive regions, initial intermolecular velocity in the collision dynamics, orientation samplings, distances checking in virtual dynamics, additional discussions on HMM, additional examples of bimolecular reactions ($NH_3+CH_3CHO$ and $NH_3+HCHO$), all species in penicillin G anion + $H_2O$ reactions and in penicillin G anion + OH reactions

(2) Cartesian coordinates of all optimized structures.

**Code Availability**

The code in this work is available from the website:

https://github.com/QHwork/repository/tree/master

## Author Information

**Corresponding Author**

E-mail: zhenggang.lan@m.scnu.edu.cn; zhenggang.lan@gmail.com.

**Notes**

The authors declare no competing financial interest.

## Acknowledge

This work is supported by the NSFC projects (Nos. 21873112, 21933011). The authors thank the Supercomputing Center, Computer Network Information Center, CAS, and National Supercomputing Center in Shenzhen for providing computational resources.

## References

(1) Cheong, P. H. Y.; Legault, C. Y.; Um, J. M.; Çelebi-Ölçüm, N.; Houk, K. N. Quantum Mechanical Investigations of Organocatalysis: Mechanisms, Reactivities, and Selectivities. *Chem. Rev.* **2011**, *111*, 5042–5137.

(2) Dewyer, A. L.; Argüelles, A. J.; Zimmerman, P. M. Methods for exploring reaction space in molecular systems. *WIREs Comput. Mol. Sci.* **2018**, *8*, e1354.

(3) Unsleber, J. P.; Reiher, M. The Exploration of Chemical Reaction Networks. *Annu. Rev. Phys. Chem.* **2020**, *71*, 121–142.

(4) Sameera, W. M. C.; Maeda, S.; Morokuma, K. Computational Catalysis Using the Artificial Force Induced Reaction Method. *Acc. Chem. Res.* **2016**, *49*, 763–773.

(5) Tentscher, P. R.; Lee, M.; von Gunten, U. Micropollutant Oxidation Studied by Quantum Chemical Computations: Methodology and Applications to Thermodynamics, Kinetics, and Reaction Mechanisms. *Acc. Chem. Res.* **2019**, *52*, 605–614.

(6) Tratnyek, P. G.; Bylaska, E. J.; Weber, E. J. In Silico Environmental Chemical Science: Properties and Processes from Statistical and Computational Modelling. *Environ. Sci.: Processes Impacts* **2017**, *19*, 188–202.

(7) Kovacevic, G.; Sabljic, A. Atmospheric Oxidation of Halogenated Aromatics: Comparative Analysis of Reaction Mechanisms and Reaction Kinetics. *Environ. Sci.:*

*Processes Impacts* **2017**, *19*, 357–369.

(8) Cai, Y. Computational Methods in Environmental and Resource Economics. *Annu. Rev. Resour. Econ.* **2019**, *11*, 59–82.

(9) Gao, J.; Truhlar, D. G. Quantum Mechanical Methods for Enzyme Kinetics. *Annu. Rev. Phys. Chem.* **2002**, *53*, 467–505.

(10) Senn, H. M.; Thiel, W. QM/MM Methods for Biomolecular Systems. *Angew. Chem. Int. Ed.* **2009**, *48*, 1198–1229.

(11) Jonsson, H. M., G.; Jacobsen, K. W. Nudged Elastic Band Method for Finding Minimum Energy Paths of Transitions in book *"Classical and Quantum Dynamics in Condensed Phase Simulations"*. World Scientific Publishing: June **1998**.

(12) Peters, B.; Heyden, A.; Bell, A. T.; Chakraborty, A. A Growing String Method for Determining Transition States: Comparison to the Nudged Elastic Band and String Methods. *J. Chem. Phys.* **2004**, *120*, 7877–7886.

(13) Goodrow, A.; Bell, A. T.; Head-Gordon, M. Transition State-Finding Strategies for Use with the Growing String Method. *J. Chem. Phys.* **2009**, *130*, 244108.

(14) E, W.; Ren, W.; Vanden-Eijnden, E. Finite Temperature String Method for the Study of Rare Events. *J. Phys. Chem. B* **2005**, *109*, 6688–6693.

(15) Quapp, W. Reaction Pathways and Projection Operators: Application to String Methods. *J. Comput. Chem.* **2004**, *25*, 1277–1285.

(16) Bergonzo, C.; Simmerling, C. An Overview of String-Based Path Sampling Methods. *Annual Reports in Computational Chemistry* **2011**, *7*, 89–97.

(17) Jessica, G. F.; Kelly, H. R.; Victor, S. B. Search for Catalysts by Inverse Design: Artificial Intelligence, Mountain Climbers, and Alchemists. *Chem. Rev.* **2019**, *119*, 6595–6612.

(18) van de Vijver, R.; Vandewiele, N. M.; Bhoorasingh, P. L.; Slakman, B. L.; Khanshan, F. S.; Carstensen, H. H.; Reyniers, M. F.; Marin, G. B.; West, R. H.; van Geem, K. M. Automatic Mechanism and Kinetic Model Generation for Gas- And Solution-Phase Processes: A Perspective on Best Practices, Recent Advances, and Future Challenges. *Int. J. Chem. Kinet.* **2015**, *47*, 199–231.

(19) Broadbelt, L. J.; Stark, S. M.; Klein, M. T. Computer Generated Pyrolysis Modeling: On-the-Fly Generation of Species, Reactions, and Rates. *Ind. Eng. Chem. Res.* **1994**, *33*, 790–799.

(20) Rappoport, D.; Galvin, C. J.; Zubarev, D. Y.; Aspuru-Guzik, A. Complex Chemical Reaction Networks from Heuristics-Aided Quantum Chemistry. *J. Chem. Theory Comput.* **2014**, *10*, 897–907.

(21) Gao, C. W.; Allen, J. W.; Green, W. H.; West, R. H. Reaction Mechanism Generator: Automatic Construction of Chemical Kinetic Mechanisms. *Comput. Phys. Commun.* **2016**, *203*, 212–225.

(22) Coley, C. W.; Barzilay, R.; Jaakkola, T. S.; Green, W. H.; Jensen, K. F. Prediction of Organic Reaction Outcomes Using Machine Learning. *ACS Cent. Sci.* **2017**, *3*, 434–443.

(23) Gómez-Bombarelli, R.; Wei, J. N.; Duvenaud, D.; Hernández-Lobato, J. M.; Sánchez-

Lengeling, B.; Sheberla, D.; Aguilera-Iparraguirre, J.; Hirzel, T. D.; Adams, R. P.; Aspuru-Guzik, A. Automatic Chemical Design Using a Data-Driven Continuous Representation of Molecules. *ACS Cent. Sci.* **2018**, *4*, 268–276.

(24) Rappoport, D.; Aspuru-Guzik, A. Predicting Feasible Organic Reaction Pathways Using Heuristically Aided Quantum Chemistry. *J. Chem. Theory Comput.* **2019**, *15*, 4099–4112.

(25) Coley, C. W.; Green, W. H.; Jensen, K. F. Machine Learning in Computer-Aided Synthesis Planning. *Acc. Chem. Res.* **2018**, *51*, 1281–1289.

(26) Zimmerman, P. M. Automated Discovery of Chemically Reasonable Elementary Reaction Steps. *J. Comput. Chem.* **2013**, *34*, 1385–1392.

(27) Suleimanov, Y. V.; Green, W. H. Automated Discovery of Elementary Chemical Reaction Steps Using Freezing String and Berny Optimization Methods. *J. Chem. Theory Comput.* **2015**, *11*, 4248–4259.

(28) Kim, Y.; Choi, S.; Kim, W. Y. Efficient Basin-Hopping Sampling of Reaction Intermediates through Molecular Fragmentation and Graph Theory. *J. Chem. Theory Comput.* **2014**, *10*, 2419–2426.

(29) Bergeler, M.; Simm, G. N.; Proppe, J.; Reiher, M. Heuristics-Guided Exploration of Reaction Mechanisms. *J. Chem. Theory Comput.* **2015**, *11*, 5712–5722.

(30) Yang, M.; Zou, J.; Wang, G.; Li, S. Automatic Reaction Pathway Search via Combined Molecular Dynamics and Coordinate Driving Method. *J. Phys. Chem. A* **2017**, *121*, 1351–1361.

(31) Yang, M.; Yang, L.; Wang, G.; Zhou, Y.; Xie, D.; Li, S. Combined Molecular Dynamics and Coordinate Driving Method for Automatic Reaction Pathway Search of Reactions in Solution. *J. Chem. Theory Comput.* **2018**, *14*, 5787–5796.

(32) Ensing, B.; de Vivo, M.; Liu, Z.; Moore, P.; Klein, M. L. Metadynamics as a Tool for Exploring Free Energy Landscapes of Chemical Reactions. *Acc. Chem. Res.* **2006**, *39*, 73–81.

(33) Shang, C.; Liu, Z. P. Stochastic Surface Walking Method for Structure Prediction and Pathway Searching. *J. Chem. Theory Comput.* **2013**, *9*, 1838–1845.

(34) Zheng, S.; Pfaendtner, J. Car-Parrinello Molecular Dynamics + Metadynamics Study of High-Temperature Methanol Oxidation Reactions Using Generic Collective Variables. *J. Phys. Chem. C* **2014**, *118*, 10764–10770.

(35) Zhang, X. J.; Liu, Z. P. Reaction Sampling and Reactivity Prediction Using the Stochastic Surface Walking Method. *Phys. Chem. Chem. Phys.* **2015**, *17*, 2757–2769.

(36) Huang, S. D.; Shang, C.; Zhang, X. J.; Liu, Z. P. Material Discovery by Combining Stochastic Surface Walking Global Optimization with a Neural Network. *Chem. Sci.* **2017**, *8*, 6327–6337.

(37) Grimme, S. Exploration of Chemical Compound, Conformer, and Reaction Space with Meta-Dynamics Simulations Based on Tight-Binding Quantum Chemical Calculations. *J. Chem. Theory Comput.* **2019**, *15*, 2847–2862.

(38) Ohno, K.; Maeda, S. A Scaled Hypersphere Search Method for the Topography of Reaction Pathways on the Potential Energy Surface. *Chem. Phys. Lett.* **2004**, *384*, 277–

282.

(39) Maeda, S.; Ohno, K.; Morokuma, K. Systematic Exploration of the Mechanism of Chemical Reactions: The Global Reaction Route Mapping (GRRM) Strategy Using the ADDF and AFIR Methods. *Phys. Chem. Chem. Phys.* **2013**, *15*, 3683–3701.

(40) Maeda, S.; Harabuchi, Y.; Takagi, M.; Saita, K.; Suzuki, K.; Ichino, T.; Sumiya, Y.; Sugiyama, K.; Ono, Y. Implementation and Performance of the Artificial Force Induced Reaction Method in the GRRM17 Program. *J. Comput. Chem.* **2018**, *39*, 233–251.

(41) Martínez-Núñez, E. An Automated Transition State Search Using Classical Trajectories Initialized at Multiple Minima. *Phys. Chem. Chem. Phys.* **2015**, *17*, 14912–14921.

(42) Martínez-Núñez, E. An Automated Method to Find Transition States Using Chemical Dynamics Simulations. *J. Comput. Chem.* **2015**, *36*, 222–234.

(43) Rossich Molina, E.; Salpin, J. Y.; Spezia, R.; Martínez-Núñez, E. On the Gas Phase Fragmentation of Protonated Uracil: A Statistical Perspective. *Phys. Chem. Chem. Phys.* **2016**, *18*, 14980–14990.

(44) Varela, J. A.; Vázquez, S. A.; Martínez-Núñez, E. An Automated Method to Find Reaction Mechanisms and Solve the Kinetics in Organometallic Catalysis. *Chem. Sci.* **2017**, *8*, 3843–3851.

(45) Rodríguez, A.; Rodríguez-Fernández, R.; A. Vázquez, S.; L. Barnes, G.; J. P. Stewart, J.; Martínez-Núñez, E. Tsscds2018: A Code for Automated Discovery of Chemical Reaction Mechanisms and Solving the Kinetics. *J. Comput. Chem.* **2018**, *39*, 1922–1930.

(46) Vázquez, S. A.; Otero, X. L.; Martínez-Núñez, E. A Trajectory-Based Method to Explore Reaction Mechanisms. *Molecules* **2018**, *23*, 3156.

(47) Jara-Toro, R. A.; Pino, G. A.; Glowacki, D. R.; Shannon, R. J.; Martínez-Núñez, E. Enhancing Automated Reaction Discovery with Boxed Molecular Dynamics in Energy Space. *Chem. Systems Chem.* **2020**, *2*, e1900024.

(48) Shannon, R. J.; Martinez-Nunez, E.; Shalashilin, D. V.; Glowacki, D. R. ChemDyME: Kinetically Steered, Automated Mechanism Generation through Combined Molecular Dynamics and Master Equation Calculations. *J. Chem. Theory Comput* **2021**, *17*, 4901-4912.

(49) Glowacki, D. R.; Paci, E.; Shalashilin, D. V. Boxed Molecular Dynamics: Decorrelation Time Scales and the Kinetic Master Equation. *J. Chem. Theory Comput.* **2011**, *7*, 1244–1252.

(50) Booth, J.; Vázquez, S.; Martínez-Núñez, E.; Marks, A.; Rodgers, J.; Glowacki, D. R.; Shalashilin, D. V. Recent Applications of Boxed Molecular Dynamics: A Simple Multiscale Technique for Atomistic Simulations. *Phil. Trans. R. Soc. A.* **2014**, *372*, 20130384.

(51) Shannon, R. J.; Amabilino, S.; O'Connor, M.; Shalishilin, D. V.; Glowacki, D. R. Adaptively Accelerating Reactive Molecular Dynamics Using Boxed Molecular Dynamics in Energy Space. *J. Chem. Theory Comput.* **2018**, *14*, 4541–4552.

(52) Martínez-Núñez, E.; Barnes, G. L.; Glowacki, D. R.; Kopec, S.; Pelaez, D.; Rodriguez, A.; Rodriguez-Fernandez, R.; Shannon, R. J.; Stewart, J. J. P.; Tahoces, P. G.; Vazquez,

S. A. AutoMeKin2021: An open-source program for automated reaction discovery. *J. Comput. Chem.* **2021**, *42*, 2036-2048.

(53) Wang, L. P.; Titov, A.; McGibbon, R.; Liu, F.; Pande, V. S.; Martínez, T. J. Discovering Chemistry with an Ab Initio Nanoreactor. *Nat. Chem.* **2014**, *6*, 1044–1048.

(54) Wang, L. P.; McGibbon, R. T.; Pande, V. S.; Martínez, T. J. Automated Discovery and Refinement of Reactive Molecular Dynamics Pathways. *J. Chem. Theory Comput.* **2016**, *12*, 638–649.

(55) Daniel, J.; Martin, J. H. *Speech and Language Processing, An Introduction to Natural Language Processing, Computational Linguistics, and Speech Recognition*. Prentice Hall: **2008**.

(56) Elstner, M.; Porezag, D.; Jungnickel, G.; Elsner, J.; Haugk, M.; Frauenheim, T.; et al. Self-consistent-charge density-functional tight-binding method for simulations of complex materials properties. *Phys. Rev. B* **1998**, *58*, 7260–7268.

(57) Seifert, G. Tight-Binding Density Functional Theory: An Approximate Kohn-Sham DFT Scheme. *J. Phys. Chem. A* **2007**, *111*, 5609–5613.

(58) Lei, T.; Guo, W.; Liu, Q.; Jiao, H.; Cao, D. B.; Teng, B.; Li, Y. W.; Liu, X.; Wen, X. D. Mechanism of Graphene Formation via Detonation Synthesis: A DFTB Nanoreactor Approach. *J. Chem. Theory Comput.* **2019**, *15*, 3654–3665.

(59) van Duin, A. C. T.; Dasgupta, S.; Lorant, F.; Goddard, W. A. ReaxFF: A reactive force field for hydrocarbons. *J. Phys. Chem. A* **2001**, *105*, 9396–9409.

(60) Chenoweth, K.; van Duin, A. C. T.; Goddard, W. A. ReaxFF reactive force field for molecular dynamics simulations of hydrocarbon oxidation. *J. Phys. Chem. A* **2008**, *112*, 1040–1053.

(61) Senftle, T.; Hong, S.; Islam, M.; et al. The ReaxFF reactive force-field: development, applications and future directions. *npj. Comput. Mater.* **2016**, *2*, 15011.

(62) Zeng, J.; Cao, L.; Xu, M.; Zhu, T.; Zhang, J. Z. H. Complex Reaction Processes in Combustion Unraveled by Neural Network-Based Molecular Dynamics Simulation. *Nat. Commun.* **2020**, *11*, 5713.

(63) Zeng, J.; Cao, L.; Chin, C. H.; Ren, H.; Zhang, J. Z. H.; Zhu, T. ReacNetGenerator: An Automatic Reaction Network Generator for Reactive Molecular Dynamics Simulations. *Phys. Chem. Chem. Phys.* **2020**, *22*, 683–691.

(64) Henkelman, G.; Uberuaga, B. P.; Jónsson, H. Climbing Image Nudged Elastic Band Method for Finding Saddle Points and Minimum Energy Paths. *J. Chem. Phys.* **2000**, *113*, 9901–9904.

(65) Fukui, K. Formulation of the reaction coordinate. *J. Phys. Chem.* **1970**, *74*, 4161-4163.

(66) Gonzalez, C.; Schlegel, H. B. Reaction Path Following in Mass-Weighted Internal Coordinates. *J. Phys. Chem.* **1990**, *94*, 5523-5527.

(67) Elmolla, E. S.; Chaudhuri, M. Comparison of Different Advanced Oxidation Processes for Treatment of Antibiotic Aqueous Solution. *Desalination* **2010**, *256*, 43–47.

(68) Watkinson, A. J.; Murby, E. J.; Costanzo, S. D. Removal of Antibiotics in Conventional and Advanced Wastewater Treatment: Implications for Environmental Discharge and Wastewater Recycling. *Water Res.* **2007**, *41*, 4164–4176.

(69) Le-Minh, N.; Khan, S. J.; Drewes, J. E.; Stuetz, R. M. Fate of Antibiotics during Municipal Water Recycling Treatment Processes. *Water Res.* **2010**, *44*, 4295–4323.

(70) Klavarioti, M.; Mantzavinos, D.; Kassinos, D. Removal of Residual Pharmaceuticals from Aqueous Systems by Advanced Oxidation Processes. *Environ. Int.* **2009**, *35*, 402–417.

(71) Isariebel, Q. P.; Carine, J. L.; Ulises-Javier, J. H.; Anne-Marie, W.; Henri, D. Sonolysis of Levodopa and Paracetamol in Aqueous Solutions. *Ultrason. Sonochem.* **2009**, *16*, 610–616.

(72) Dias, I. N.; Souza, B. S.; Pereira, J. H. O. S.; Moreira, F. C.; Dezotti, M.; Boaventura, R. A. R.; Vilar, V. J. P. Enhancement of the Photo-Fenton Reaction at near Neutral PH through the Use of Ferrioxalate Complexes: A Case Study on Trimethoprim and Sulfamethoxazole Antibiotics Removal from Aqueous Solutions. *Chem. Eng. J.* **2014**, *247*, 302–313.

(73) Fukahori, S.; Fujiwara, T. Modeling of Sulfonamide Antibiotic Removal by $TiO_2$/High-Silica Zeolite HSZ-385 Composite. *J. Hazard. Mater.* **2014**, *272*, 1–9.

(74) Zazouli, M. A.; Susanto, H.; Nasseri, S.; Ulbricht, M. Influences of Solution Chemistry and Polymeric Natural Organic Matter on the Removal of Aquatic Pharmaceutical Residuals by Nanofiltration. *Water Res.* **2009**, *43*, 3270–3280.

(75) Kamranifar, M.; Al-Musawi, T. J.; Amarzadeh, M.; Hosseinzadeh, A.; Nasseh, N.; Qutob, M.; Arghavan, F. S. Quick Adsorption Followed by Lengthy Photodegradation Using $FeNi_3$@$SiO_2$@ZnO: A Promising Method for Complete Removal of penicillin G from Wastewater. *J. Water Process Eng.* **2021**, *40*, 101940.

(76) Chavoshan, S.; Khodadadi, M.; Nasseh, N. Photocatalytic Degradation of penicillin G from Simulated Wastewater Using the UV/ZnO Process: Isotherm and Kinetic Study. *J. Environ. Health Sci. Engineer.* **2020**, *18*, 107–117.

(77) Zimmerman, P. M. Navigating molecular space for reaction mechanisms: an efficient, automated procedure. *Mol. Simul.* **2015**, *41*, 43-54.

(78) Glendening, E. D.; Landis, C. R.; Weinhold, F. Natural bond orbital methods. *WIREs Comput. Mol. Sci.* **2012**, *2*, 1–42.

(79) Oláh, J.; van Alsenoy, C.; Sannigrahi, A. B. Condensed Fukui Functions Derived from Stockholder Charges: Assessment of Their Performance as Local Reactivity Descriptors. *J. Phys. Chem. A* **2002**, *106*, 3885–3890.

(80) Parr, R. G.; Yang, W. Density Functional Approach to the Frontier-Electron Theory of Chemical Reactivity. *J. Am. Chem. Soc.* **1984**, *106*, 4049–4050.

(81) Yang, W.; Mortier, W. J. The Use of Global and Local Molecular Parameters for the Analysis of the Gas-Phase Basicity of Amines. *J. Am. Chem. Soc.* **1986**, *108*, 5708–5711.

(82) Wigner, E. On the Quantum Correction For Thermodynamic Equilibrium. *Phys. Rev.* **1932**, *40*, 749–759.

(83) Proctor, E. A.; Ding, F.; Dokholyan, N. V. Discrete molecular dynamics. *WIREs Comput. Mol. Sci.* **2011**, *1*, 80-92.

(84) Haynes, W. M. *CRC Handbook of Chemistry and Physics*. CRC Press: **2014**.

(85) Gaussian 16, Revision B.01, M. J. Frisch, G. W. Trucks, H. B. Schlegel, G. E. Scuseria, M. A. Robb, J. R. Cheeseman, G. Scalmani, V. Barone, G. A. Petersson, H. Nakatsuji, X. Li, M. Caricato, A. V. Marenich, J. Bloino, B. G. Janesko, R. Gomperts, B. Mennucci, H. P. Hratchian, J. V. Ortiz, A. F. Izmaylov, J. L. Sonnenberg, D. Williams-Young, F. Ding, F. Lipparini, F. Egidi, J. Goings, B. Peng, A. Petrone, T. Henderson, D. Ranasinghe, V. G. Zakrzewski, J. Gao, N. Rega, G. Zheng, W. Liang, M. Hada, M. Ehara, K. Toyota, R. Fukuda, J. Hasegawa, M. Ishida, T. Nakajima, Y. Honda, O. Kitao, H. Nakai, T. Vreven, K. Throssell, J. A. Montgomery, Jr., J. E. Peralta, F. Ogliaro, M. J. Bearpark, J. J. Heyd, E. N. Brothers, K. N. Kudin, V. N. Staroverov, T. A. Keith, R. Kobayashi, J. Normand, K. Raghavachari, A. P. Rendell, J. C. Burant, S. S. Iyengar, J. Tomasi, M. Cossi, J. M. Millam, M. Klene, C. Adamo, R. Cammi, J. W. Ochterski, R. L. Martin, K. Morokuma, O. Farkas, J. B. Foresman, and D. J. Fox, Gaussian, Inc., Wallingford CT, **2016**.

(86) NBO Version 3.1, E. D. Glendening, A. E. Reed, J. E. Carpenter, and F. Weinhold.

(87) Becke, A. D. A new mixing of Hartree–Fock and local density-functional theories. *J. Chem. Phys.* **1993**, *98*, 1372-1377.

(88) Lee, C.; Yang, W.; Parr, R. G. Development of the Colle-Salvetti correlation-energy formula into a functional of the electron density. *Phys. Rev. B* **1988**, *37*, 785-789.

(89) Ditchfield, R.; Hehre, W. J.; Pople, J. A. Self-Consistent Molecular-Orbital Methods. IX. An Extended Gaussian-Type Basis for Molecular-Orbital Studies of Organic Molecules. *J. Chem. Phys.* **1971**, *54*, 724-728.

(90) Hehre, W. J.; Ditchfield, R.; Pople, J. A. Self—Consistent Molecular Orbital Methods. XII. Further Extensions of Gaussian—Type Basis Sets for Use in Molecular Orbital Studies of Organic Molecules. *J. Chem. Phys.* **1972**, *56*, 2257-2261.

(91) Hariharan, P. C.; Pople, J. A. The Influence of polarization functions on molecular-orbital hydrogenation energies. *Theor. Chem. Acc.* **1973**, *28*, 213-22.

(92) Gordon, M. S. The isomers of silacyclopropane. *Chem. Phys. Lett.* **1980**, *76*, 163-68.

(93) Francl, M. M.; Pietro, W. J.; Hehre, W. J.; Binkley, J. S.; Gordon, M. S.; DeFrees, D. J.; Pople, J. A. Self-consistent molecular orbital methods. XXIII. A polarization-type basis set for second-row elements. *J. Chem. Phys.* **1982**, *77*, 3654-3665.

(94) Petersson, G. A.; Bennett, A.; Tensfeldt, T. G.; Al-Laham, M. A.; Shirley, W. A.; Mantzaris, J. A complete basis set model chemistry. I. The total energies of closed-shell atoms and hydrides of the first-row elements. *J. Chem. Phys.* **1988**, *89*, 2193-2218.

(95) Binning, Jr. R. C.; Curtiss, L. A. Compact contracted basis-sets for 3rd-row atoms - GA-KR. *J. Comp. Chem.* **1990**, *11*, 1206-16.

(96) Blaudeau, J.-P.; McGrath, M. P.; Curtiss, L. A.; Radom, L. Extension of Gaussian-2 (G2) theory to molecules containing third-row atoms K and Ca. *J. Chem. Phys.* **1997**, *107*, 5016-5021.

(97) Rassolov, V. A.; Ratner, M. A.; Pople, J. A.; Redfern, P. C.; Curtiss, L. A. 6-31G* Basis Set for Third-Row Atoms. *J. Comp. Chem.* **2001**, *22*, 976-84.

(98) Hariharan, P. C.; Pople, J. A. Accuracy of $AH_n$ equilibrium geometries by single determinant molecular orbital theory. *Mol. Phys.* **1974**, *27*, 209-214.

(99) Rassolov, V. A.; Pople, J. A.; Ratner, M. A.; Windus, T. L. 6-31G* basis set for atoms K through Zn. *J. Chem. Phys.* **1998**, *109*, 1223-29.

(100) Grimme, S.; Antony, J.; Ehrlich, S.; Krieg, H. A consistent and accurate ab initio parametrization of density functional dispersion correction (DFT-D) for the 94 elements H-Pu. *J. Chem. Phys.* **2010**, *132*, 154104.

(101) Lu, T.; Chen, F. Multiwfn: A Multifunctional Wavefunction Analyzer. *J. Comput. Chem.* **2012**, *33*, 580–592.

(102) Binkley, J. S.; Pople, J. A.; Hehre, W. J. Self-Consistent Molecular Orbital Methods. 21. Small Split-Valence Basis Sets for First-Row Elements. *J. Am. Chem. Soc.* **1980**, *102*, 939-47.

(103) Gordon, M. S.; Binkley, J. S.; Pople, J. A.; Pietro, W. J.; Hehre, W. J. Self-Consistent Molecular Orbital Methods. 22. Small Split-Valence Basis Sets for Second-Row Elements. *J. Am. Chem. Soc.* **1982**, *104*, 2797-803.

(104) Pietro, W. J.; Francl, M. M.; Hehre, W. J.; Defrees, D. J.; Pople, J. A.; Binkley, J. S. Self-Consistent Molecular Orbital Methods. 24. Supplemented small split-valence basis-sets for 2nd-row elements. *J. Am. Chem. Soc.* **1982**, *104*, 5039-48.

(105) Dobbs, K. D.; Hehre, W. J. Molecular-orbital theory of the properties of inorganic and organometallic compounds. 4. Extended basis-sets for 3rd row and 4th row, main-group elements. *J. Comp. Chem.* **1986**, *7*, 359-78.

(106) Dobbs, K. D.; Hehre, W. J. Molecular-orbital theory of the properties of inorganic and organometallic compounds. 5. Extended basis-sets for 1st-row transition-metals. *J. Comp. Chem.* **1987**, *8*, 861-79.

(107) Dobbs, K. D.; Hehre, W. J. Molecular-orbital theory of the properties of inorganic and organometallic compounds. 6. Extended basis-sets for 2nd-row transition-metals. *J. Comp. Chem.* **1987**, *8*, 880-93.

(108) Neese, F. The ORCA Program System. *WIREs Comput. Mol. Sci.* **2012**, *2*, 73–78.

# Supporting Information for

# An automatic approach to explore multi-reaction mechanism for medium-sized bimolecular reactions via collision dynamics simulations and transition state searches

*Qinghai Cui[1], Jiawei Peng[1], Chao Xu[1,2], Zhenggang Lan[1,*]*

[1] Guangdong Provincial Key Laboratory of Chemical Pollution and Environmental Safety and MOE Key Laboratory of Environmental Theoretical Chemistry, SCNU Environmental Research Institute, School of Environment, South China Normal University, Guangzhou 510006, P. R. China.

[2] Key Laboratory of Theoretical Chemistry of Environment, Ministry of Education; School of Chemistry, South China Normal University, Guangzhou 510006, P. R. China.

* Corresponding Author

E-mail: zhenggang.lan@m.scnu.edu.cn; zhenggang.lan@gmail.com

# 1. Optimized structure of penicillin G anion

Figure S1. Optimized structure of penicillin G anion with atomic labels.

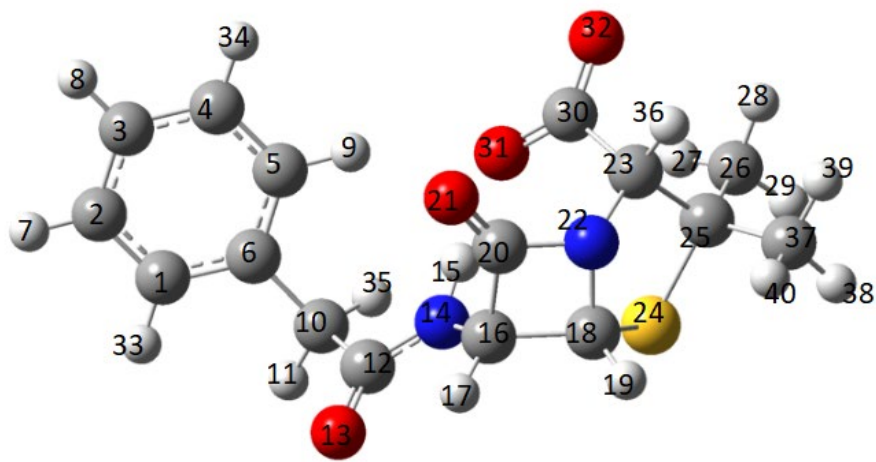

## 2. NBO charges and CFF values of penicillin G anion

Table S1. NBO Charges and CFF Values of Penicillin G anion

| Atom | NBO | CFF ($f_+$) | CFF ($f_-$) | CFF ($f_{radical}$) |
| --- | --- | --- | --- | --- |
| 1(C) | -0.112 | 0.073 | 0.015 | 0.044 |
| 2(C) | -0.088 | 0.074 | 0.014 | 0.044 |
| 3(C) | -0.088 | 0.074 | 0.015 | 0.044 |
| 4(C) | -0.095 | 0.073 | 0.014 | 0.043 |
| 5(C) | -0.122 | 0.072 | 0.017 | 0.045 |
| 6(C) | 0.113 | 0.072 | 0.015 | 0.043 |
| 7(H) | 0.055 | 0.012 | 0.001 | 0.007 |
| 8(H) | 0.056 | 0.012 | 0.001 | 0.007 |
| 9(H) | 0.0148 | 0.008 | 0.003 | 0.006 |
| 10(C) | -0.330 | 0.019 | 0.009 | 0.014 |
| 11(H) | 0.097 | 0.006 | 0.001 | 0.003 |
| 12(C) | 0.570 | 0.042 | 0.014 | 0.028 |
| 13(O) | -0.547 | 0.025 | 0.066 | 0.046 |
| 14(N) | -0.493 | 0.020 | 0.038 | 0.029 |
| 15(H) | 0.331 | 0.003 | 0.007 | 0.005 |
| 16(C) | -0.029 | 0.017 | 0.017 | 0.017 |
| 17(H) | 0.116 | 0.005 | 0.003 | 0.004 |
| 18(C) | -0.077 | 0.030 | 0.015 | 0.023 |
| 19(H) | 0.111 | 0.008 | 0.006 | 0.007 |
| 20(C) | 0.562 | 0.057 | 0.018 | 0.037 |
| 21(O) | -0.471 | 0.039 | 0.050 | 0.044 |
| 22(N) | -0.453 | 0.026 | 0.044 | 0.035 |
| 23(C) | 0.032 | 0.014 | 0.035 | 0.025 |
| 24(S) | 0.047 | 0.091 | 0.088 | 0.089 |
| 25(C) | -0.127 | 0.031 | 0.009 | 0.020 |
| 26(C) | -0.303 | 0.009 | 0.004 | 0.006 |
| 27(H) | 0.144 | 0.003 | 0.001 | 0.002 |
| 28(H) | 0.121 | 0.003 | 0.001 | 0.002 |
| 29(H) | 0.073 | 0.005 | 0.001 | 0.003 |
| 30(C) | 0.580 | 0.009 | 0.050 | 0.029 |
| 31(O) | -0.604 | 0.006 | 0.183 | 0.094 |
| 32(O) | -0.594 | 0.006 | 0.225 | 0.116 |
| 33(H) | 0.066 | 0.011 | 0.001 | 0.006 |
| 34(H) | 0.083 | 0.010 | 0.001 | 0.005 |
| 35(H) | 0.122 | 0.007 | 0.002 | 0.005 |
| 36(H) | 0.100 | 0.005 | 0.007 | 0.006 |
| 37(C) | -0.282 | 0.010 | 0.005 | 0.008 |
| 38(H) | 0.092 | 0.005 | 0.001 | 0.003 |
| 39(H) | 0.093 | 0.005 | 0.001 | 0.003 |
| 40(H) | 0.107 | 0.004 | 0.001 | 0.002 |

## 3. Selection of reactive regions

In the selection of reactive regions, it should be possible to allow that the positive charged atom only interacts with the negative charged atom. However, this may result in too many chosen reactive pairs.

In practice, the selection was performed according to the following procedure.

In the calculations of the penicillin G anion + water reactions, the whole water was taken as a reactant region because of its small size. The reactive region of the penicillin G anion was defined by the below procedure:

(1) The NBO charge and CFF value for each atom were obtained by electronic-structure calculations.

(2) The NBO charges were first considered. We first defined a fragment composed of three adjacent atoms. All H atoms were not included in these fragments because they generally display very small NBO charges. The "effective NBO value" of this fragment was given by summing the absolute NBO charges of these three atoms. Please notice that such effective NBO value is not the overall NBO charge of this fragment. And then, we tried to pick up 10 fragments displaying the largest effective NBO values.

(3) We tried to calculate the effective CFF value of all fragments again. If the chosen fragments selected in Step (2) lay in the top 50% of the list of effective CFF values, this fragment was taken as the reactive fragment.

(4) After the selection of the reactive fragments, we added back the H atoms connecting to these fragments. This gave us the reactive regions.

(5) When two reactive regions shared the same atom, we may decide whether to merge them or not by considering their connectivity. When both regions belonged to the same conjugated ring unit, we simply merged two regions together to define a larger reactive region. Otherwise, we did not merge them. For the region in which the sum of the CFF values of all atoms is smaller, we kept it unchanged to define the first reactive region. For the region show the larger total CFF value, we deleted the shared atom that belongs to the first region, and this defines the second reactive region.

In the calculations of the penicillin G anion + OH radical reactions, the whole OH radical was taken as the whole reactant region. Here the OH radical in principle shows the very high reactivity and it can react with many parts of penicillin G anion. Therefore, we selected five reactive regions for penicillin G anion only according to the CFF values.

At the end, in the selected reactive region, the NBO charges and CFF values of the included atoms are given in Table S2. These values can only serve as a rough estimation of the overall reactivity of the whole reactive regions. Please notice that these values are fully determined by the whole reactive region, not individual atoms. Thus, other individual atoms out of the reaction regions may also show the similar NBO charges and CFF values.

In the treatment of reactions involve only small compounds ($NH_3$, HCHO, $CH_3CHO$), we did not perform such pre-screening step by using NBO charges and CFF values.

Certainly, the above selection standard may not always be optimal. We may directly use some loose threshold values to include more atoms. We even can try to use different criteria to examine the convergence to check whether the observed low-energy reactions do not change with such parameters. At the same time, the involvement of chemical intuition may be highly helpful.

Table S2. NBO and CFF Values in the Chosen Reactive Region (without H atom)

| System | NBO | CFF |
|---|---|---|
| Penicillin G anion + $H_2O$ | Positive > 0.4<br>Negative < -0.4 | 0.018 |
| Penicillin G anion + OH radical | —— | 0.02 |

## 4. Initial intermolecular velocity in the collision dynamics

For two species, we tried to connect the two centers of reactive regions by a line. Then we added the inter-molecular velocities along the direction of this line and the initial velocities were determined by the chosen collision energies. This makes sure that the direct collision takes place between two reactive regions. In the initial sampling, the total linear momentum was set to zero. Figure S2 shows the motion of a water molecule towards the selected reactive center of a penicillin G anion.

Figure S2. The collision dynamics takes place between two reactants, i.e., a water molecule and the selected reactive region of a penicillin G anion. (a) shows the initial geometries and their inter-molecular velocities. (b) shows the relative motion of the water molecule along the trajectory before collision. The velocities of the two fragments were determined by the collision energy within the constrain of setting the total linear momentum to be zero.

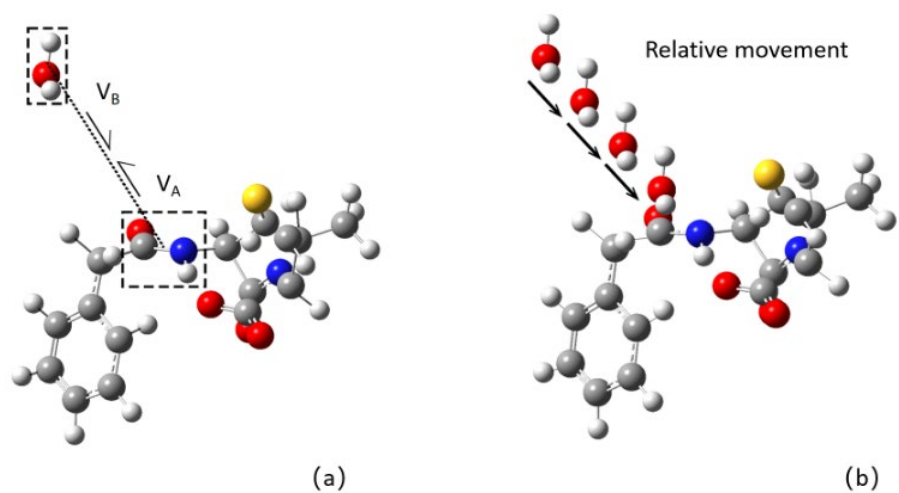

## 5. Orientation samplings

The sampling of different orientations of two reactive species is explained by considering a typical system: a $NH_3$ molecule and a HCHO molecule. Here the $NH_3$ molecule was located at the center of a sphere with radius 7 Å. Next a HCHO molecule was generated at a randomly chosen point on the surface of this sphere. The orientation of this HCHO molecule was also randomly sampled.

Similar to the approach explained in Figure S2, we performed the collision dynamics with proper sampling of the initial intermolecular velocity. Figure S3 shows the sampling of all possible orientations.

Different initial intermolecular velocities were taken in account in real BOMD collision dynamics. We also considered the sampling of the vibrational motion by using Wigner sampling of the lowest vibrational state.

In the virtual dynamics, the velocity of the larger compound was set to be zero and the velocity of the smaller compound was taken according to the low boundary of the collision energy. No sampling based on the vibrational motion was considered.

Figure S3. Overlap of initial sampling conditions with different orientations. Here the example gives the NH$_3$ and HCHO reactions. (a) shows all 700 starting conditions and (b) shows all 500 starting conditions. The arrows show the chosen HCHO with different positions and orientations.

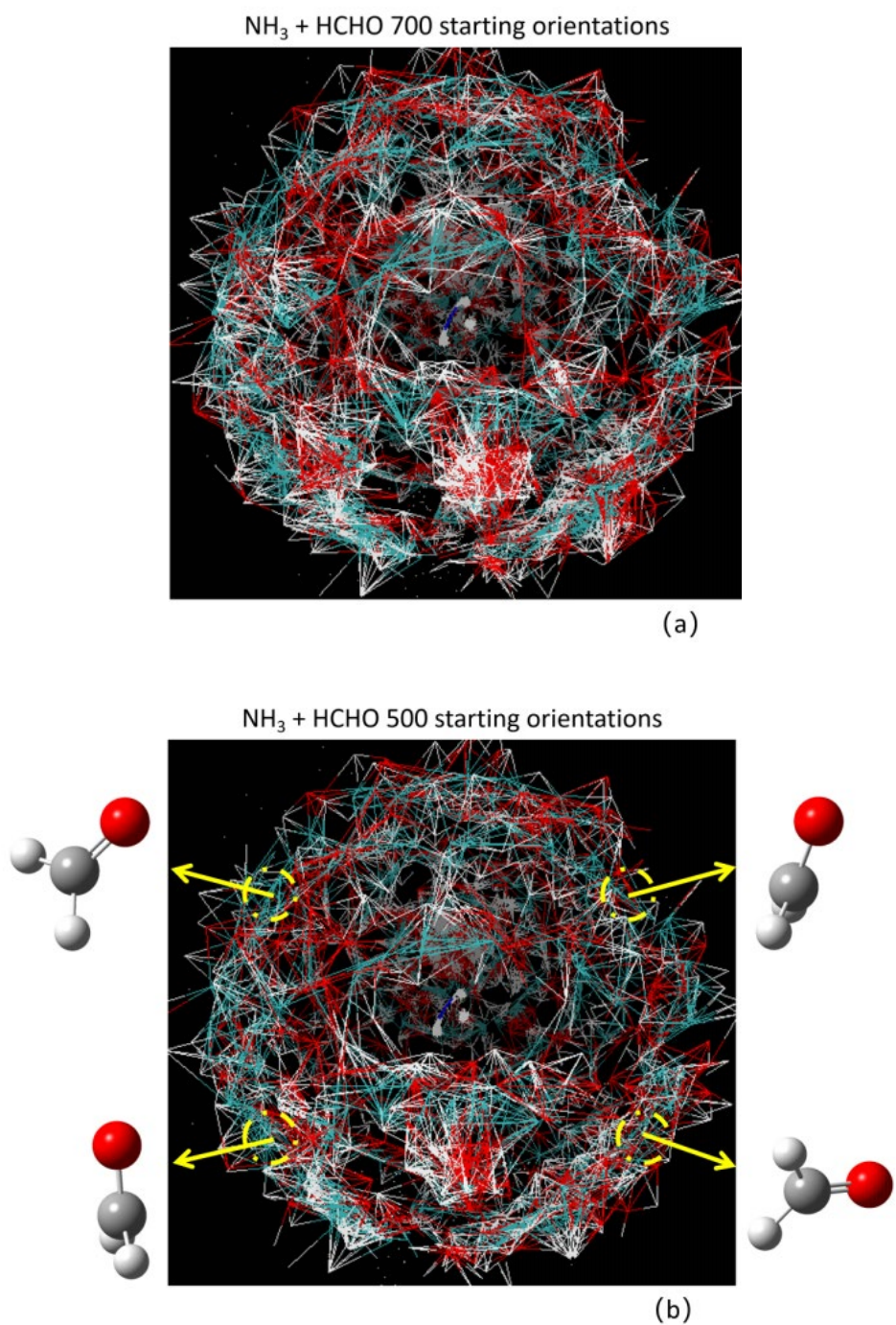

## 6. Distances checking in virtual dynamics

In the virtual dynamics, we monitored all pairs of inter-atomic distances between two species, as shown in Figure S4. Suppose that Atom_a, Atom_b and Atom_c belong to MOLE_B (for instance H or O), the reaction region of MOLE_A and non-reaction region of MOLE_A, respectively. When the distance between Atom_a and Atom_b is lower than the 1.4 times of their covalent bond length, we assume that the possible collision may take place between two reaction regions. Such collision is assumed to be effective, only when the distance between Atom_a and Atom_c is always larger than 1.4 times of their corresponding covalent bond length before the collision between Atom_a and Atom_b.

Figure S4. All inter-atomic distances between two reaction centers are monitored in the virtual dynamics.

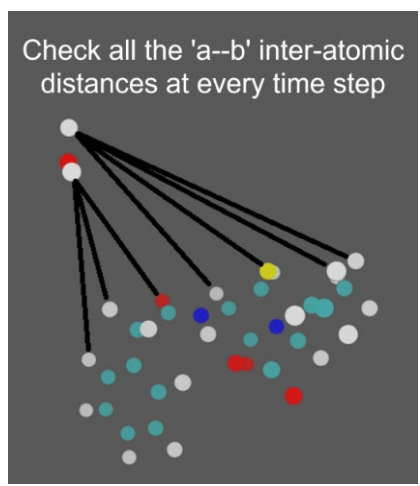

# 7. Additional discussions on HMM

In the HMM model, we define a hidden state (not visible) that evolves over time being and an observable state that is visible at any given time. We wish to derive the "hidden state" status at a given time step by minoring "observable state". Here a hidden state does not correspond to only a single observable state. Instead, under such hidden state, the appearance of the visible observable is determined by a probability.

In the current work, the breaking and formation of a chemical bond were characterized by the connectivity matrix. Next, we defined the binary hidden states based on the changing of connectivity matrix. When any element of the connectivity matrix changes (bond breaking or formation), the observable is set to be 1. If all elements remain unchanged (no bond breaking or formation) between two adjacent steps, the observable is set to be 0. Such binary observable states give us the direct indicator on the changing of inter-atomic connectivity. However, the changing of the connectivity matrix may reflect the bond breaking or formation, while it may only indicate the large vibrational motion. At the same time, in principle, the breaking of an old chemical bond or the formation of a new bond may prefer to give the observable value of 1 but not always. Instead, it gives either 0 or 1 according to some possibility. The HMM model simply takes this possibility into account.

In the current BOMD, the energy is very high and thus the very strong stretching motion may take place. This gives too many oscillations of the binary observable states between 0 and 1. If we take all of them into the next optimization step, the computational cost becomes extremely high. At the same time, the selection of all state jumps is also not necessary because many of them just correspond to the strong bond stretching motion. We used HMM to filter out many oscillations and only chose a few representative ones. In one word, the original trajectory gives a large number of oscillations of binary state, while the employment of the HMM model largely reduces the number of the selection events in the BOMD. Similar approach was also used by previous works to reduce the number of the chosen reaction events.[1,2,3]

## 8. Additional examples of bimolecular reactions

We tried to test this theoretical method by examining bimolecular reactions of small systems given by the works by Zimmerman.[4,5] Here we chose two test systems:

(1) a $NH_3$ molecule + a $CH_3CHO$ molecule;

(2) a $NH_3$ molecule + a $HCHO$ molecule.

Since all these compounds are very small, the procedure would reckon the whole molecules as reactive regions and it is not necessary to perform the prescreening of the virtual dynamics. Here only the test BOMD was run with different collision energies (range 0.37 - 0.66 Hartree), and finally, we selected 0.48 Hartree as the lowest value to run the formal BOMD simulation. The employment of HMM model and reaction pathway construction followed the similar approaches.

### (a) NH₃ + CH₃CHO reactions:

Figure S5. All NH₃ + CH₃CHO reactions found by the current approach. E is the electronic energy barrier of the elementary step.

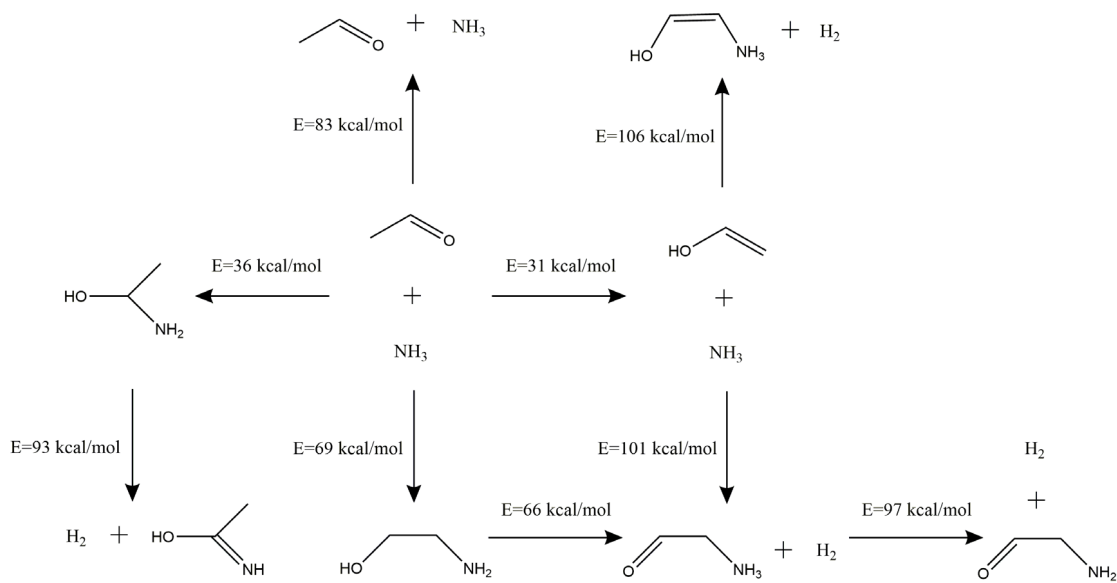

Figure S6. The two lowest-energy-barrier NH₃ + CH₃CHO reactions and the relevant transition state structures.

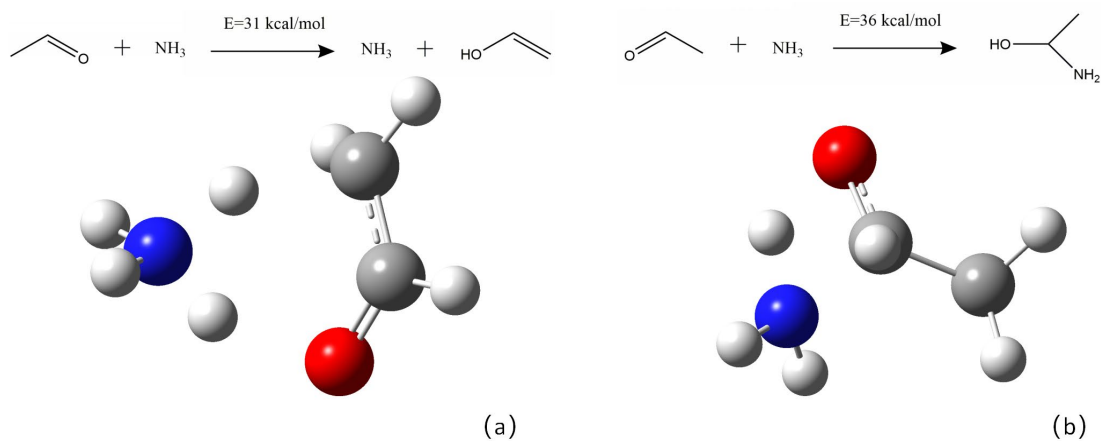

As shown in Figure S5 and Figure S6, several reactions are found by using the current approach. Two lowest reactions (Figure S6 (a) and (b)) are same as the ones shown in the work by Zimmerman.[4] Except them, we found several high-energy reactions.

Under low collision energy (0.48 Hartree), we found one low-energy N-H addition reaction (36 kcal/mol) and a few of high-energy reactions (66 kcal/mol, 69 kcal/mol, 83 kcal/mol, 93 kcal/mol, 97 kcal/mol).

Under medium collision energy (0.57 Hartree), the N-H addition reaction is also found and a new low-energy keto-enol tautomerisation reaction (31 kcal/mol) appeared. And a few high-energy reactions (66 kcal/mol, 69 kcal/mol, 93 kcal/mol, 97 kcal/mol, 101 kcal/mol, 106 kcal/mol) are found.

Under high collision energy (0.66 Hartree), we located only one low-energy reaction, the same low-energy N-H addition reaction, and a few high-energy reactions (66 kcal/mol, 69 kcal/mol, 93 kcal/mol, 97 kcal/mol).

Therefore, we believe the calculation is converged and it is not necessary to improve the collision energy anymore. All results on the number of the found TS structures *v.s.* the number of trajectories are shown in Table S3.

Table S3. Number of TS Structures *v.s.* Number of Trajectories for the $NH_3$ + $CH_3CHO$ Reactions

| Collision Energy (Hartree) | Number of Trajectories | TS Numbers (different connectivity) | Reaction Barriers (kcal/mol) |
|---|---|---|---|
| 0.48 | 500 | 6 | 36 66,69,83,93,97 |
| 0.57 | 500 | 8 | 31,36 66,69,93,97,101,106 |
| 0.66 | 500 | 5 | 36 66,69,93,97 |

**(b) NH₃+HCHO reactions:**

Figure S7. All NH$_3$ + HCHO reactions found by the current approach. E is the energy barrier of the elementary step.

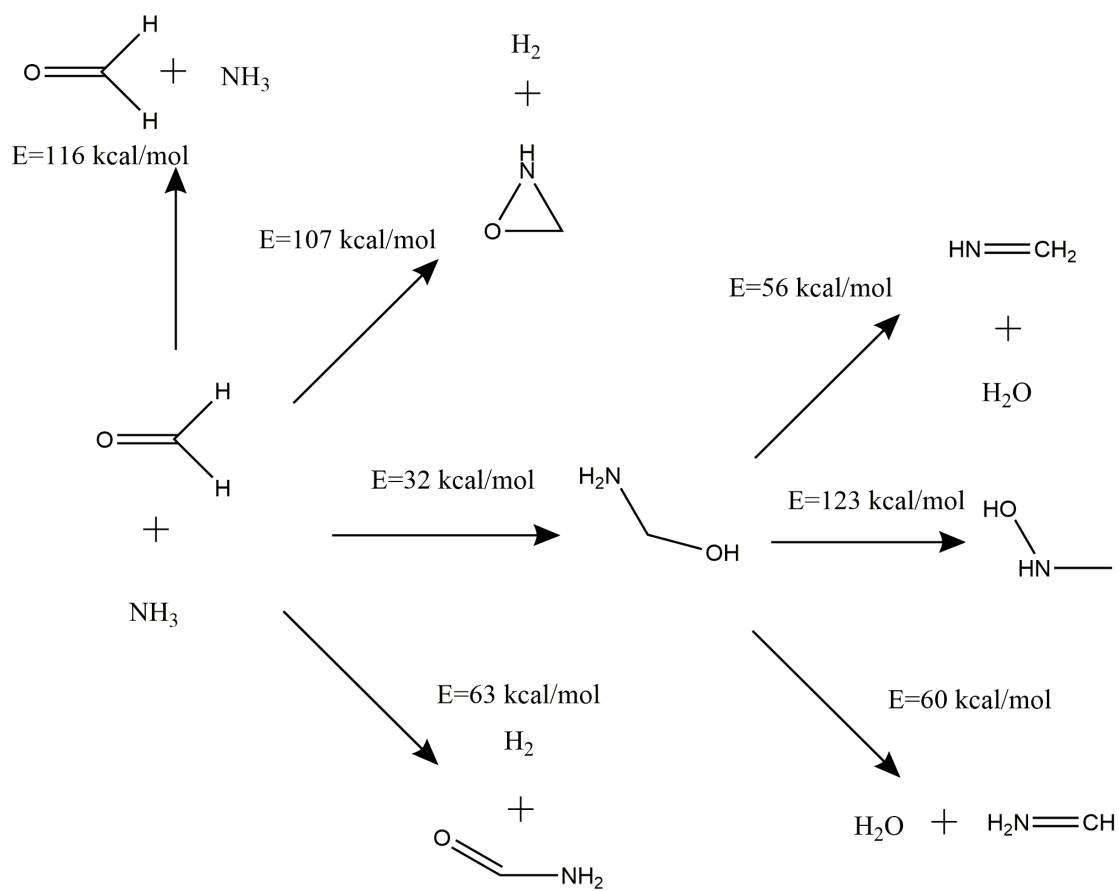

Figure S8. Two lowest-energy NH$_3$ + HCHO reactions and the relevant transition state structures.

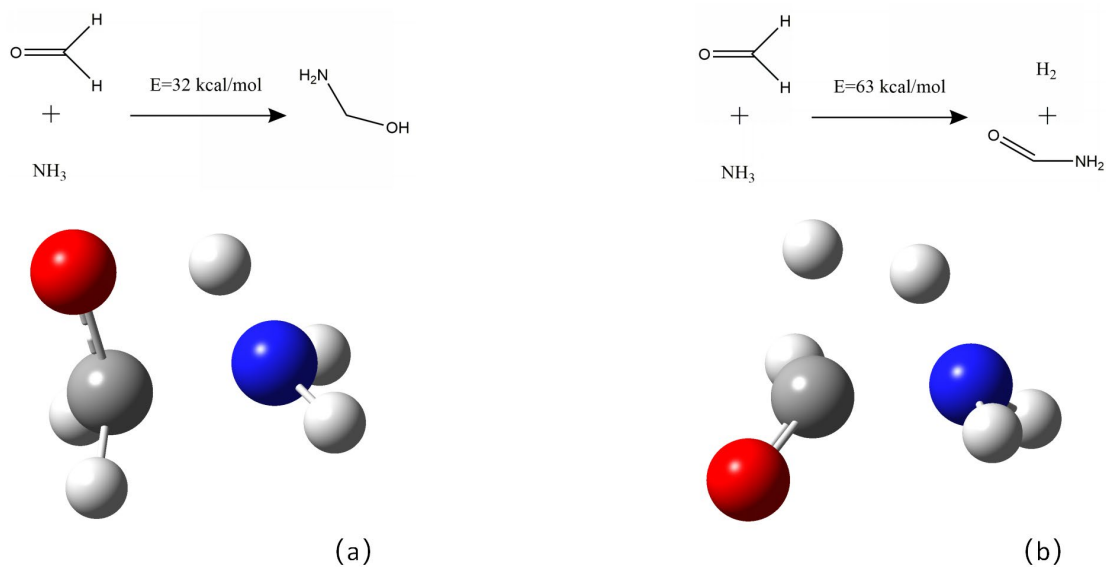

As shown in Figure S7 and Figure S8, several reaction pathways are found by using the current approach.

Among them, the lowest reaction pathway (Figure S8 (a)) is the same as the one shown in the work of Zimmerman[5].

The second lowest-energy reaction (Figure S8 (b)) is a little different to their result. Although both works clarified that this is a H$_2$ elimination reaction, the details are different. The reactive barrier is 63 kcal/mol in our work and 80 kcal/mol in the work by Zimmerman. The detailed study on the origin of such difference is beyond the scope of our current investigation, because the barriers of both reactions are larger than 40 kcal/mol. The products are H$_2$, CO and NH$_3$ in the work by Zimmerman, while in our work they are H$_2$ and NH$_2$CO.

Except them, we also found several high-energy reactions.

To summarize our observations on these two examples, we more-or-less got the consistent results as the findings by the work of Zimmerman for the low-energy (< 40 kcal/mol) reactions. For the higher-energy reactions (for example, the second lowest NH3 + HCHO reaction with the barrier height larger than 40 kcal/mol), some deviations exist between the work of Zimmerman and the current one. For the very high-energy reactions, two approaches may give different answers. However, we only focus on the low-energy reactions, and thus we believe the current approach is trustworthy.

The finding of all reactions is in fact a NP-hard problem. In this sense, all approaches just tried to find the reactions as many as possible and no available method can make sure that all reactions are found.

# 9. All species in penicillin G anion + H₂O reactions

Figure S9. All species involved in the penicillin G anion + $H_2O$ reaction network. Here we label each structure according to the following rule: the capital letters, **R**, **TS** and **P** that denote reactants, TS, and products respectively. The numbers after these capital letters correspond to the reaction indices given in Figure 4. The dashed line and the blue arrows show the key atomic distances with the unit of Å.

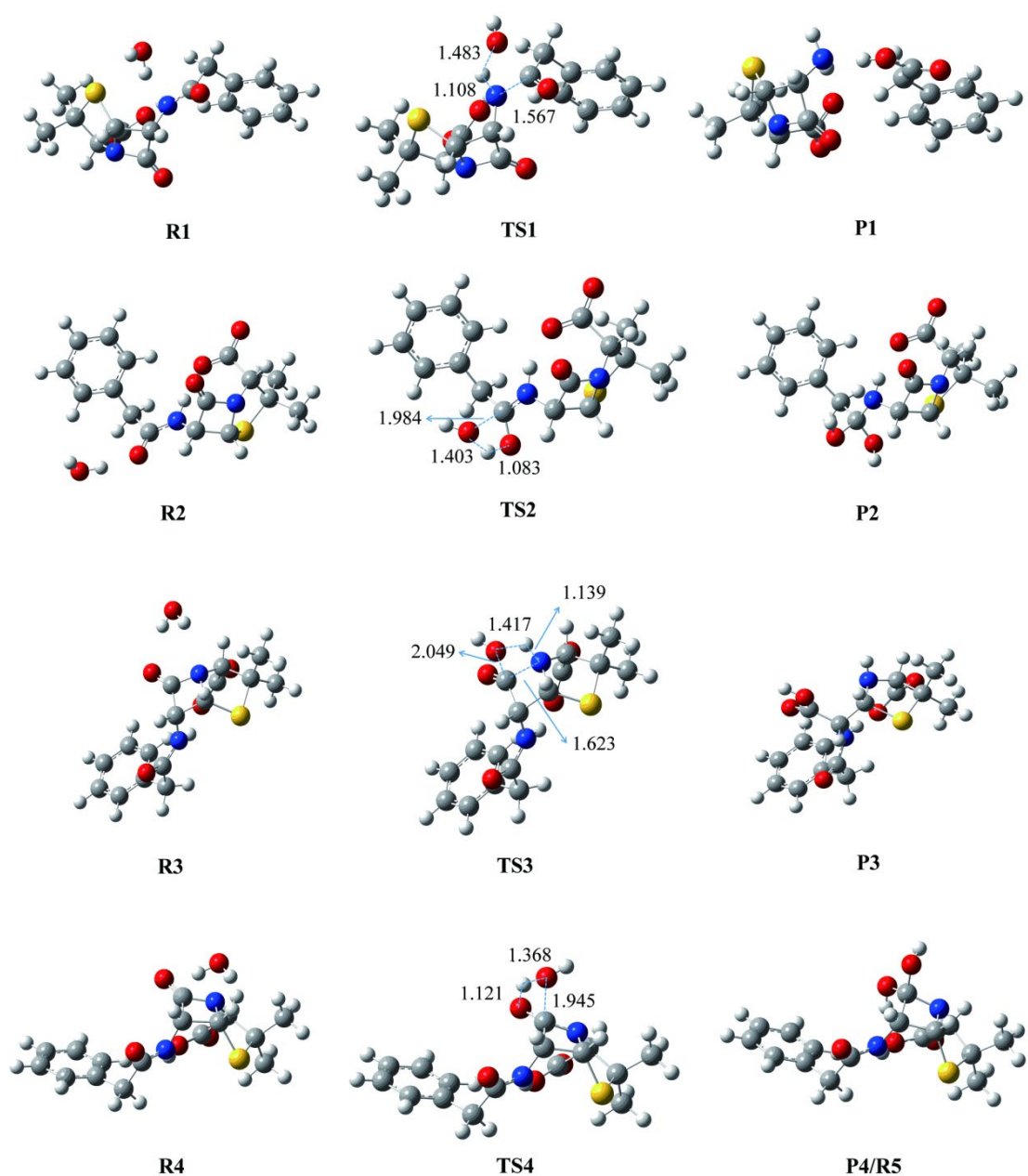

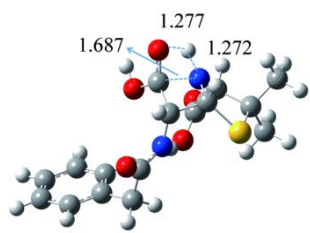
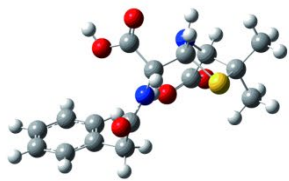
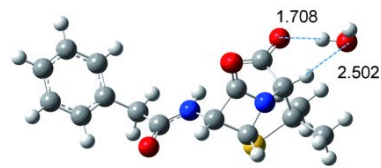

**TS5**  **P5**  **R6**

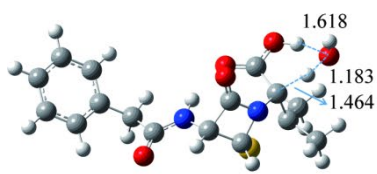
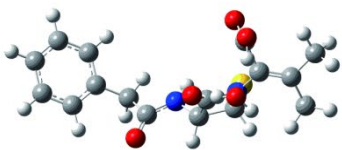
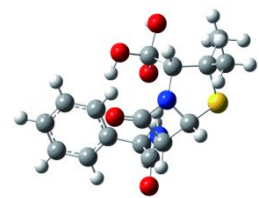

**TS6**  **P6**  **R7**

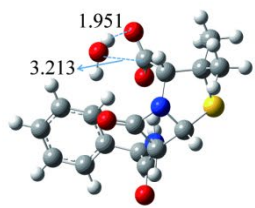
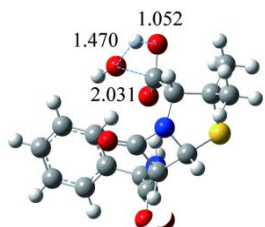
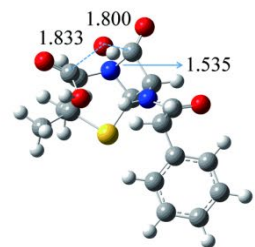

**TS7**  **P7/R8/R9**  **TS8**

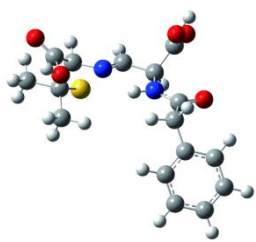
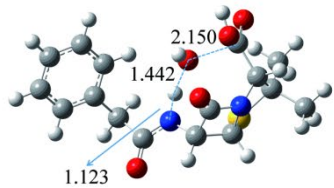
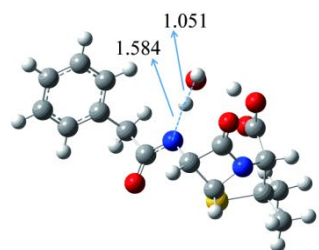

**P8**  **TS9**  **P9**

## 10. All species in penicillin G anion + OH reactions

Figure S10. All species involved in the penicillin G anion + OH reaction network. Here we label each structure according to the following rule: the capital letters, **R**, **TS** and **P** that denote reactants, TS, and products respectively. The numbers after these capital letters correspond to the reaction indices given in Figure 7.

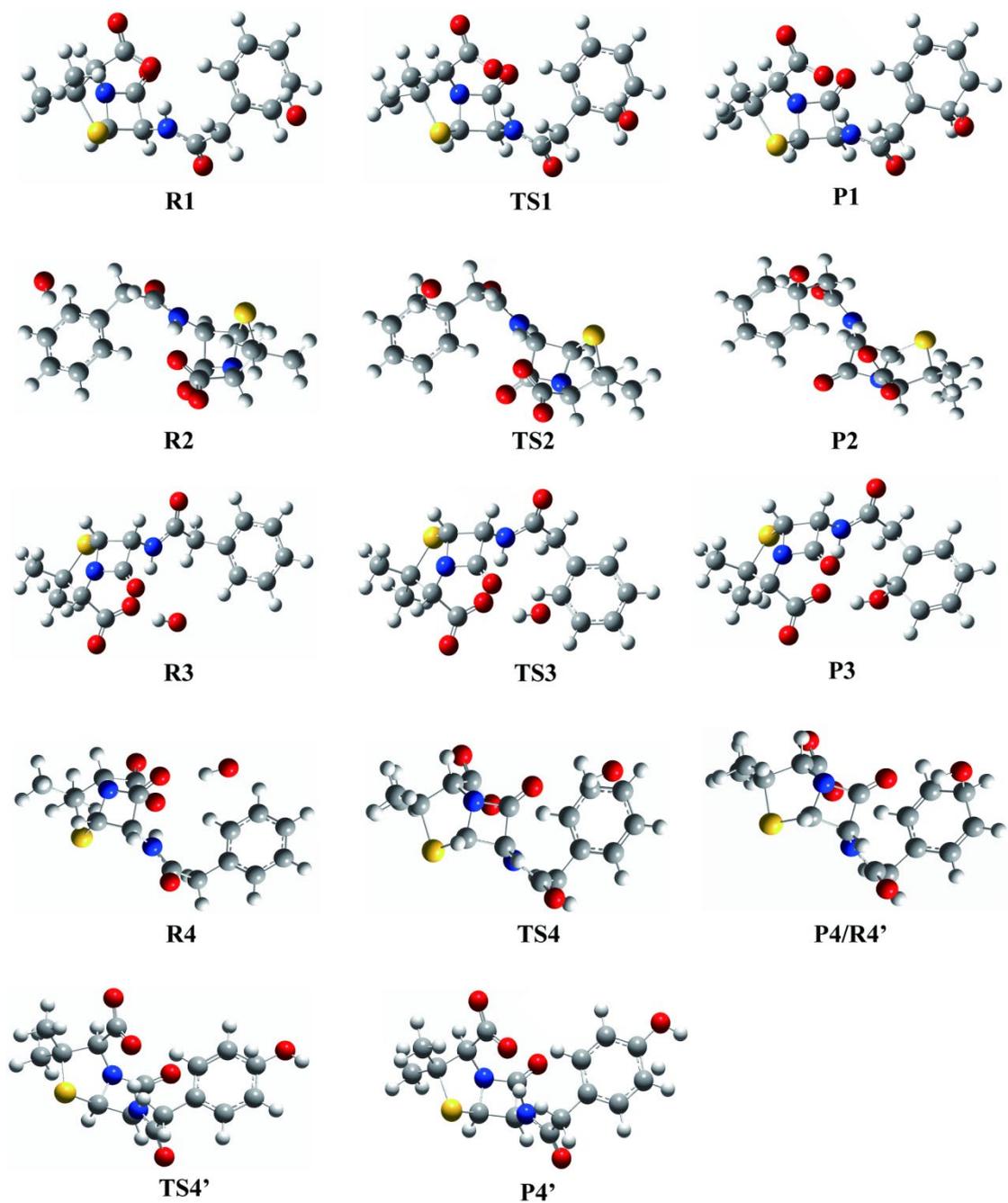

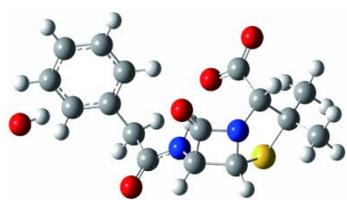
**R5**

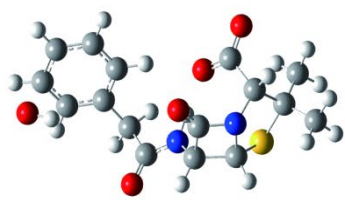
**TS5**

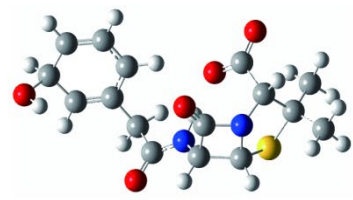
**P5/R5'**

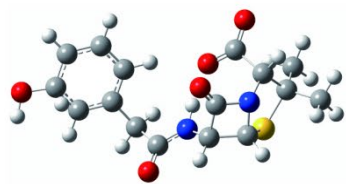
**TS5'**

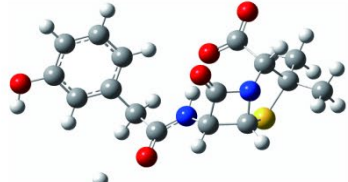
**P5'**

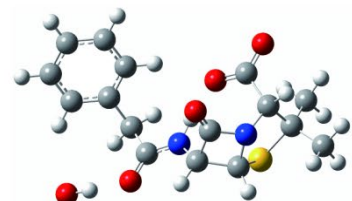
**R6**

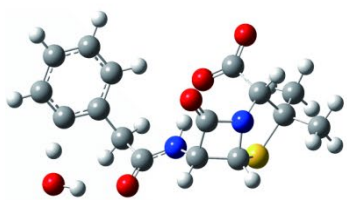
**TS6**

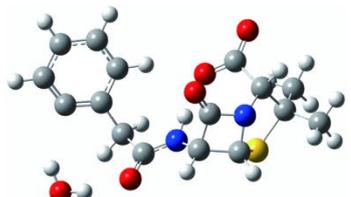
**P6**

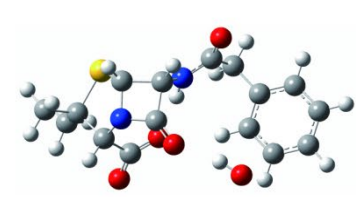
**R7**

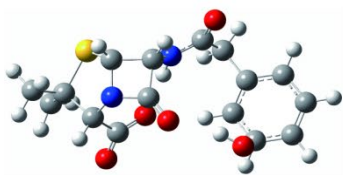
**TS7**

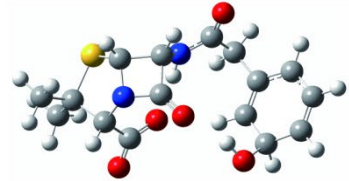
**P7/R7'**

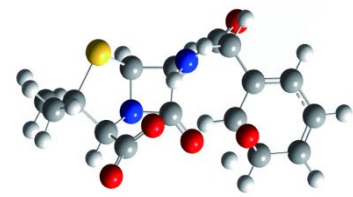
**TS7'**

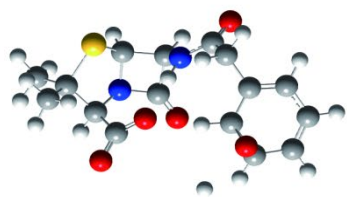
**P7'**

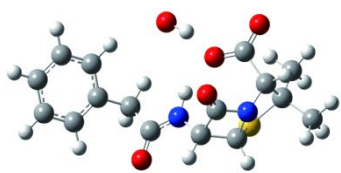
**R8**

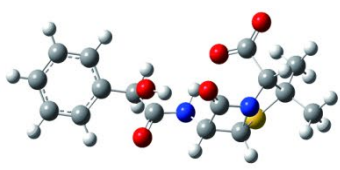
**TS8**

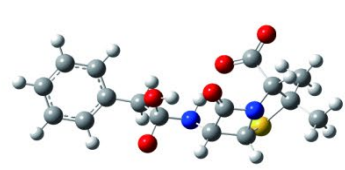
**P8/R8'**

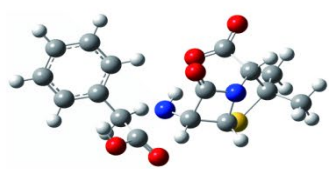
**TS8'**

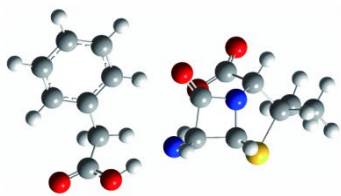
**P8'**

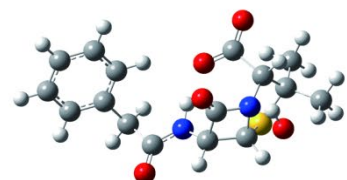
**R9**

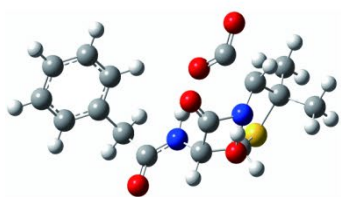
**TS9**

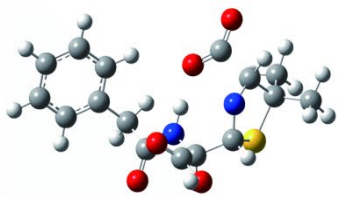
**P9**

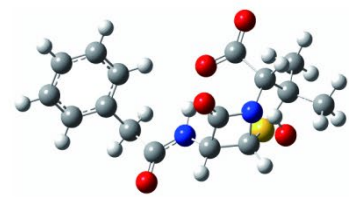
**R10**

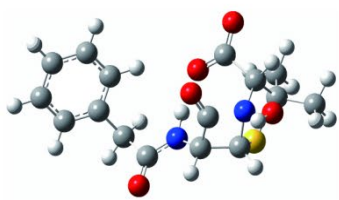
**TS10**

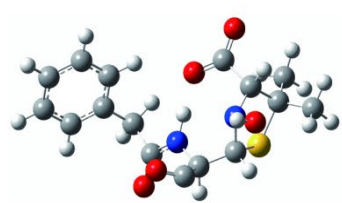
**P10**

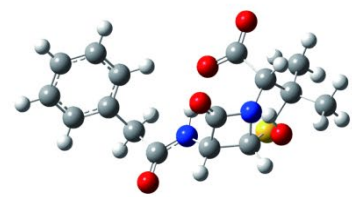
**R11**

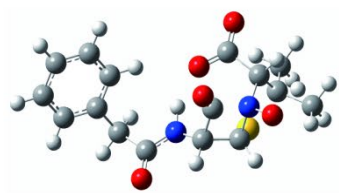
**TS11**

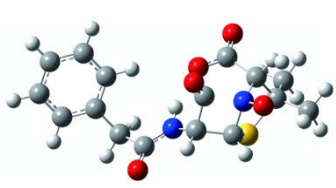
**P11**

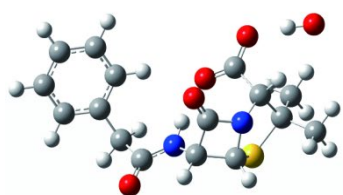 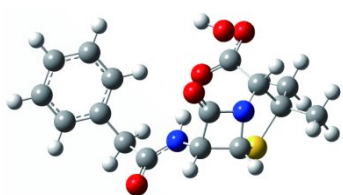 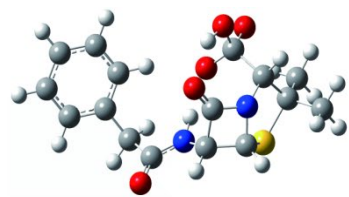

**R12**     **TS12**     **P12**

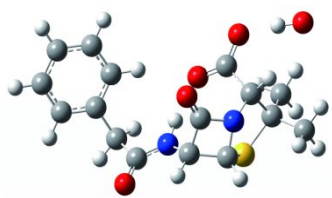 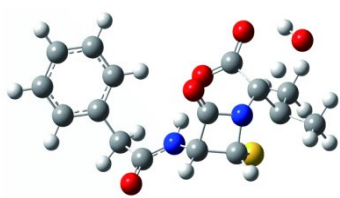 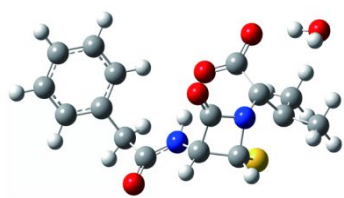

**R13**     **TS13**     **P13**

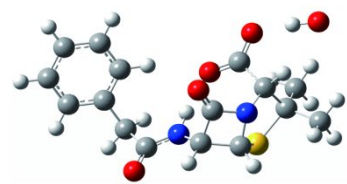 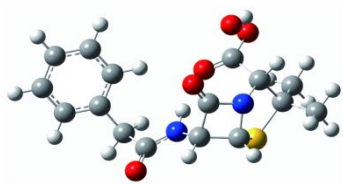 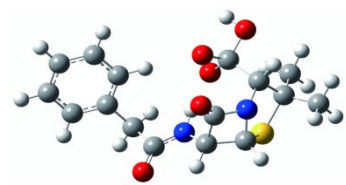

**R14**     **TS14**     **P14**

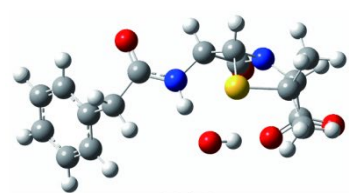 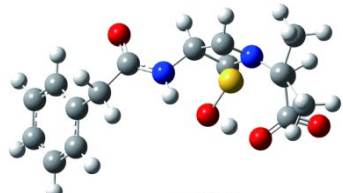 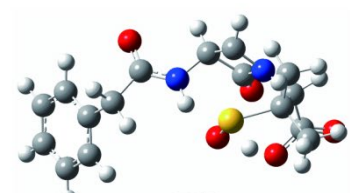

**R15**     **TS15**     **P15**

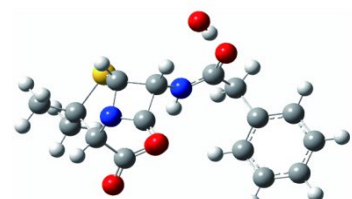 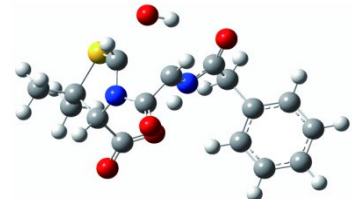 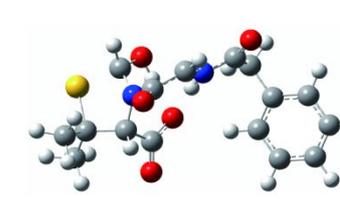

**R16**     **TS16**     **P16**

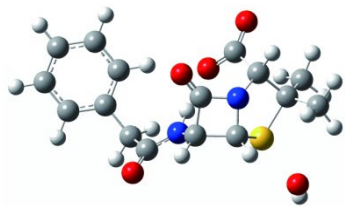
**R17**

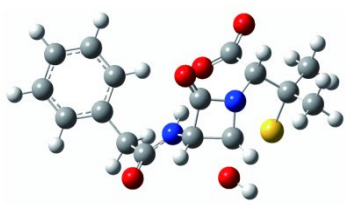
**TS17**

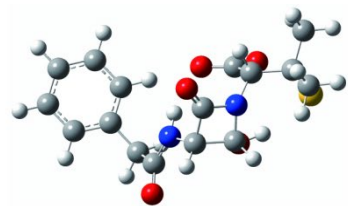
**P17**

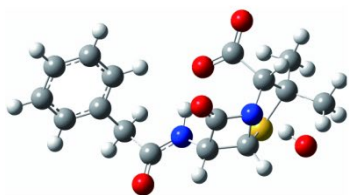
**R18**

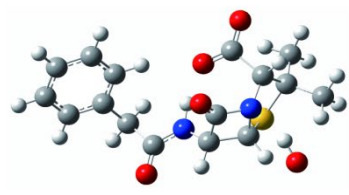
**TS18**

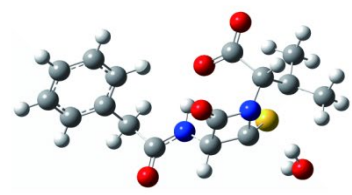
**P18/R18'/R19'**

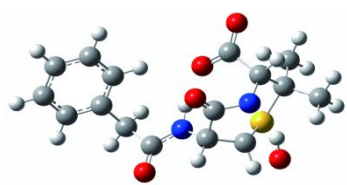
**TS18'**

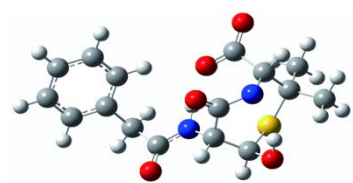
**P18'**

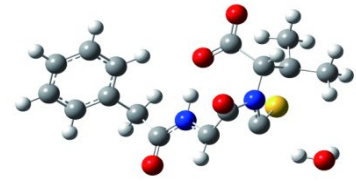
**TS19'**

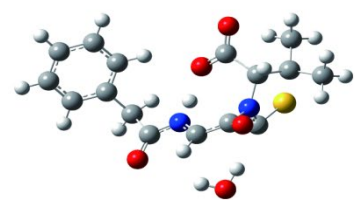
**P19'**

# References

[1] Wang, L. P.; Titov, A.; McGibbon, R.; Liu, F.; Pande, V. S.; Martínez, T. J. Discovering Chemistry with an Ab Initio Nanoreactor. *Nat. Chem.* **2014**, *6*, 1044-1048.

[2] Wang, L. P.; McGibbon, R. T.; Pande, V. S.; Martínez, T. J. Automated Discovery and Refinement of Reactive Molecular Dynamics Pathways. *J. Chem. Theory Comput.* **2016**, *12*, 638–649.

[3] Zeng, J.; Cao, L.; Chin, C. H.; Ren, H.; Zhang, J. Z. H.; Zhu, T. ReacNetGenerator: An Automatic Reaction Network Generator for Reactive Molecular Dynamics Simulations. *Phys. Chem. Chem. Phys.* **2020**, *22*, 683–691.

[4] Zimmerman, P. M. Navigating Molecular Space for Reaction Mechanisms: An Efficient, Automated Procedure. *Mol. Simulat.* **2015**, *41*, 43-54.

[5] Zimmerman, P. M. Automated Discovery of Chemically Reasonable Elementary Reaction Steps. *J. Comput. Chem.* **2013**, *34*, 1385-1392.